\pdfoutput=1
\documentclass[runningheads]{llncs}
\usepackage[T1]{fontenc}
\usepackage{graphicx}
\usepackage{amsmath, amssymb, stmaryrd, mathtools}
\usepackage{proof}
\usepackage[table]{xcolor}
\usepackage{tikz}
\usepackage{todonotes}
\usepackage{cite}
\usepackage[appendix=append]{apxproof}
\usetikzlibrary{cd, math, calc, decorations.pathmorphing}
\usetikzlibrary{automata, backgrounds, trees, hobby, positioning, automata, arrows}
\usepackage{ifthen}
\usepackage{subcaption}
\usepackage{booktabs, multirow}
\usepackage{makecell}
\usepackage{hyperref, cleveref}

\allowdisplaybreaks

\newcommand{\myparagraph}[1]{\noindent{\bfseries {\sffamily #1}}\quad}

\definecolor{darkgray}{rgb}{0.31,0.31,0.33}
\definecolor[named]{lipicsGray}{rgb}{0.31,0.31,0.33}
\definecolor[named]{lipicsBulletGray}{rgb}{0.60,0.60,0.61}
\definecolor[named]{lipicsLineGray}{rgb}{0.51,0.50,0.52}
\definecolor[named]{lipicsLightGray}{rgb}{0.85,0.85,0.86}
\definecolor[named]{lipicsYellow}{rgb}{0.99,0.78,0.07}

\usepackage{wrapfig}
\theoremstyle{plain}
\newtheoremrep{mylemma}{Lemma}[section]
\newtheoremrep{myproposition}[mylemma]{Proposition}
\newtheoremrep{mytheorem}[mylemma]{Theorem}
\newtheorem{mycorollary}[mylemma]{Corollary}
\theoremstyle{definition}
\newtheorem{mydefinition}[mylemma]{Definition}
\newtheorem{myassumption}[mylemma]{Assumption}
\newtheorem{myremark}[mylemma]{Remark}

\newtheorem{myexample}[mylemma]{Example}

\newtheorem{mynotation}[mylemma]{Notation}

\theoremstyle{remark}

\crefname{mydefinition}{Def.}{Def.}
\crefname{myproposition}{Prop.}{Prop.}
\crefname{mylemma}{Lem.}{Lem.}
\crefname{mytheorem}{Thm.}{Thm.}
\crefname{mycorollary}{Cor.}{Cor.}
\crefname{mynotation}{Notation}{Notation}
\crefname{myexample}{Example}{Example}
\crefname{myremark}{Remark}{Remark}
\crefname{myassumption}{Assum.}{Assum.}
\crefformat{section}{{\S}#2#1#3}
\crefname{algorithm}{Algo.}{Algo.}

\crefname{notation}{Notation}{Notation}
\crefname{figure}{Fig.}{Fig.}

\newsavebox{\pullback}
\sbox\pullback{%
    \begin{tikzpicture}%
    \draw (0,0) -- (1ex,0ex);%
    \draw (1ex,0ex) -- (1ex,1ex);%
    \end{tikzpicture}%
}
\newsavebox{\pushout}
\sbox\pushout{%
    \begin{tikzpicture}%
    \draw (-1ex,0) -- (-1ex,1ex);%
    \draw (-1ex,1ex) -- (0,1ex);%
    \end{tikzpicture}%
}

\tikzset{heavy/.style={color=lipicsYellow, line width=1.2mm, rounded corners, line cap=round}}
\tikzset{light/.style={color=lipicsGray, line width=0.3mm, rounded corners, line cap=round}}

\usepackage[ruled]{algorithm}
\usepackage[noend]{algpseudocode}
\makeatletter
\algrenewcommand\ALG@beginalgorithmic{\footnotesize}
\makeatother

\newcommand{\algorithmiccontinue}{\textbf{continue}}
\newcommand{\Continue}{\State \algorithmiccontinue}

\newcommand{\itemref}[1]{\ref{#1}}

\newcommand{\order}{\mathcal{O}}

\newcommand{\catc}{\mathbb{C}}
\newcommand{\cate}{\mathbb{E}}
\newcommand{\sets}{\mathbf{Set}}
\newcommand{\eqrel}{\mathbf{EqRel}}
\newcommand{\idfunc}{\mathrm{Id}}
\newcommand{\idmorph}{\mathrm{id}}
\newcommand{\sub}{\mathop{\mathbf{Sub}}}
\newcommand{\obj}{\mathop{\mathrm{ob}}}
\newcommand{\monoto}{\rightarrowtail}
\newcommand{\image}{\mathop{\mathrm{Im}}}
\newcommand{\powset}{\mathcal{P}}
\newcommand{\nondet}{N}

\newcommand{\distr}{\mathcal{D}}
\newcommand{\rellift}[1]{\mathrm{Rel}({#1})}

\newcommand{\blank}{{-}}

\newcommand{\clat}{\mathbf{CLat}_{\sqcap}}
\newcommand{\lift}[1]{\overline{#1}}
\newcommand{\pull}[1]{#1^{*}}
\newcommand{\push}[1]{#1_{*}}

\newcommand{\nat}{\mathbb{N}}

\newcommand{\dirty}{\mathsf{di}}
\newcommand{\clean}{\mathsf{cl}}

\newcommand{\vertex}{V}

\newcommand{\ch}{\mathop{\mathrm{ch}}\nolimits}
\newcommand{\lch}{\mathop{\mathrm{lch}}\nolimits}
\newcommand{\subtree}{\mathop{\mathrm{tr}}\nolimits}
\newcommand{\leaves}{L}
\newcommand{\treepath}{\mathop{\mathrm{path}}}
\newcommand{\lpath}{\mathop{\mathrm{lpath}}}

\newcommand{\PRNaive}{\mathsf{fPR}^{\mathrm{naive}}}
\newcommand{\PRHopcroft}{\mathsf{fPR}^{\mathrm{H}}}
\newcommand{\PRHopcroftER}{\mathsf{fPR}^{\mathrm{H}\mathchar`-\mathbf{ER}}}
\newcommand{\PRHopcroftEqRel}[1]{\mathsf{fPR}^{\mathrm{H}\text{-}\mathbf{ER}}_{#1}}
\newcommand{\wCard}{w_{\mathrm{C}}}  
\newcommand{\wPred}{w_{\mathrm{P}}}  
\newcommand{\wReach}{w_{\mathrm{R}}}  

\pgfsetlayers{background, main}
\newcommand{\convexpath}[2]{
[   
    create hullnodes/.code={
        \global\edef\namelist{#1}
        \foreach [count=\counter] \nodename in \namelist {
            \global\edef\numberofnodes{\counter}
            \node at (\nodename) [draw=none,name=hullnode\counter] {};
        }
        \node at (hullnode\numberofnodes) [name=hullnode0,draw=none] {};
        \pgfmathtruncatemacro\lastnumber{\numberofnodes+1}
        \node at (hullnode1) [name=hullnode\lastnumber,draw=none] {};
    },
    create hullnodes
]
($(hullnode1)!#2!-90:(hullnode0)$)
\foreach [
    evaluate=\currentnode as \previousnode using \currentnode-1,
    evaluate=\currentnode as \nextnode using \currentnode+1
    ] \currentnode in {1,...,\numberofnodes} {
-- ($(hullnode\currentnode)!#2!-90:(hullnode\previousnode)$)
  let \p1 = ($(hullnode\currentnode)!#2!-90:(hullnode\previousnode) - (hullnode\currentnode)$),
    \n1 = {atan2(\y1,\x1)},
    \p2 = ($(hullnode\currentnode)!#2!90:(hullnode\nextnode) - (hullnode\currentnode)$),
    \n2 = {atan2(\y2,\x2)},
    \n{delta} = {-Mod(\n1-\n2,360)}
  in 
    {arc [start angle=\n1, delta angle=\n{delta}, radius=#2]}
}
-- cycle
}
\newcommand{\hobbyconvexpath}[2]{
[   
    create hobbyhullnodes/.code={
        \global\edef\namelist{#1}
        \foreach [count=\counter] \nodename in \namelist {
            \global\edef\numberofnodes{\counter}
            \node at (\nodename)
[draw=none,name=hobbyhullnode\counter] {};
        }
        \node at (hobbyhullnode\numberofnodes)
[name=hobbyhullnode0,draw=none] {};
        \pgfmathtruncatemacro\lastnumber{\numberofnodes+1}
        \node at (hobbyhullnode1)
[name=hobbyhullnode\lastnumber,draw=none] {};
    },
    create hobbyhullnodes
]
($(hobbyhullnode1)!#2!-90:(hobbyhullnode0)$)
\pgfextra{
  \gdef\hullpath{}
\foreach [
    evaluate=\currentnode as \previousnode using int(\currentnode-1),
    evaluate=\currentnode as \nextnode using int(\currentnode+1)
    ] \currentnode in {1,...,\numberofnodes} {
    \ifnum\currentnode=1\relax
    \xdef\hullpath{([closed=true]$(hobbyhullnode\currentnode)!#2!180:(hobbyhullnode\previousnode)$)
  ..($(hobbyhullnode\nextnode)!0.5!(hobbyhullnode\currentnode)$)}
    \else
    \xdef\hullpath{\hullpath
  ..($(hobbyhullnode\currentnode)!#2!180:(hobbyhullnode\previousnode)$)
  ..($(hobbyhullnode\nextnode)!0.5!(hobbyhullnode\currentnode)$)}
    \fi
    \ifx\currentnode\numberofnodes
    \else
    \xdef\hullpath{\hullpath
  ..($(hobbyhullnode\nextnode)!#2!-90:(hobbyhullnode\currentnode)$)}
    \fi
}
}
\hullpath
}
\newcommand*{\colorsetbkwh}[1]{
    \ifthenelse{\equal{#1}{clean}}{\colorlet{NodeTextColor}{black}}{\colorlet{NodeTextColor}{white}}
    \ifthenelse{\equal{#1}{clean}}{\colorlet{NodeFillColor}{white}}{\colorlet{NodeFillColor}{black}}
}
\newcommand{\partitioningexample}[6]{
    \begin{tikzpicture}[
        baseline=-0.8cm,
        ->,  
        node distance=5cm, 
        every state/.style={thick, fill=gray!10}, 
        initial text=$ $, 
        thick,
        scale=0.32, 
        every node/.style={transform shape, font=\huge},
        every edge/.append style={transform shape},
        use Hobby shortcut
    ]
        \colorsetbkwh{#2}
        \node[state, text=NodeTextColor, fill=NodeFillColor] (x) {$x$};
        \colorsetbkwh{#3}
        \node[state, above right of=x, yshift=-2.4cm, text=NodeTextColor, fill=NodeFillColor] (y) {$y$};
        \colorsetbkwh{#5}
        \node[state, below right of=y, xshift=-2.7cm, accepting, text=NodeTextColor, fill=NodeFillColor] (w) {$w$};
        \colorsetbkwh{#4}
        \node[state, left of=w, yshift=-2.8cm, xshift=2.9cm, text=NodeTextColor, fill=NodeFillColor] (z) {$z$};
        \colorsetbkwh{#6}
        \node[state, below of=x, yshift=1cm, text=NodeTextColor, fill=NodeFillColor] (v) {$v$};
        \draw   (x) edge[above, bend left] node{$a$} (y)
        (y) edge[above, bend left=10] node{$a$} (x)
        (w) edge[above, loop right=10cm, min distance=10mm,in=-30, out=30] node{$a$} (w)
        (w) edge[above, bend left] node{$a$} (z)
        (x) edge[above] node{$b$} (z)
        (y) edge[above, below, left=0.3] node{$b$} (z)
        (x) edge[above, bend left=5] node{$b$} (w)
        (y) edge[above, bend left] node{$b$} (w)
        (v) edge[above] node{$b$} (z)
        (z) edge[above, bend left=5] node{$b$} (w)
        (v) edge[above, loop, min distance=10mm,out=60,in=120] node{$a$} (v) ;
        {#1}
    \end{tikzpicture}
}

\begin{document}
\title{Explicit Hopcroft's Trick in Categorical Partition Refinement}
%
%
\author{Takahiro Sanada\inst{1}\orcidID{0000-0003-3409-6963} \and
Ryota Kojima\inst{2} \and
Yuichi Komorida\inst{3,4}\orcidID{0000-0002-3371-524} \and
Koko Muroya\inst{2}\orcidID{0000-0003-1760-988} \and
Ichiro Hasuo\inst{3,4}\orcidID{0000-0002-8300-4650}}
\authorrunning{T. Sanada et al.}
%
\institute{
    Fukui Prefectural University, Fukui, Japan \\
    \email{tsanada@fpu.ac.jp} \and
    RIMS, Kyoto University, Japan \\
    \email{\{kojima, kmuroya\}@kurims.kyoto-u.ac.jp} \and
    National Institute of Informatics, Japan \and
    SOKENDAI, Tokyo, Japan \\
    \email{komorin@nii.ac.jp} \ \email{i.hasuo@acm.org}
}
\maketitle              
\begin{abstract}
  Algorithms for \emph{partition refinement} are actively studied for a variety of systems, often with the optimisation called \emph{Hopcroft's trick}.
  However, the low-level description of those algorithms  in the literature often obscures the essence of Hopcroft's trick. 
  Our contribution is twofold.
  Firstly, we present a novel formulation of Hopcroft's trick in terms of general trees with weights.
  This clean and explicit formulation---we call it \emph{Hopcroft's inequality}---is crucially used in our second contribution, namely a general partition refinement algorithm that is \emph{functor-generic} (i.e.\ it works for a variety of systems such as (non-)deterministic  automata and Markov chains).
  Here we build on recent works on coalgebraic partition refinement but depart from them with the use of \emph{fibrations}.
  In particular, our fibrational notion of \emph{$R$-partitioning} exposes a concrete tree structure to which Hopcroft's inequality readily applies.
  It is notable that our fibrational framework accommodates such  algorithmic analysis  on  the categorical level of abstraction.
\end{abstract}

\section{Introduction}\label{sec:intro}

\emph{Partition refinement} refers to a class of algorithms
that computes behavioural equivalence of various types of systems---such as the language equivalence for deterministic finite automata (DFAs), bisimilarity for labelled transition systems (LTSs) and Markov chains, etc.---by a fixed-point iteration.
Such algorithms also yield \emph{quotients} of state spaces,
making systems smaller and thus easier to analyse e.g.\ by model checking. 

Since its original introduction by Moore~\cite{Moore56} for DFAs,
partition refinement has been actively studied for enhanced performance and generality.
On the performance side, Hopcroft~\cite{Hopcroft71} introduced what is now called \emph{Hopcroft's trick}
that greatly improves the asymptotic complexity.
The original paper~\cite{Hopcroft71} is famously hard to crack;
works such as~\cite{Gries73, Knuutila01} present its reformulation, again focusing on DFAs.
On the generality side, partition refinement for systems other than DFAs has been pursued, such as 
LTSs \cite{KanellakisSmolka90, Valmari09},
probabilistic transition systems with non-determinism \cite{Groote+18},
weighted automata \cite{LombardySakarovitch22}, and
weighted tree automata \cite{Hogberg+07a,Hogberg+07b}.
Hopcroft's trick is used in many of these works for enhanced performance, too.

Such a variety of target systems is uniformly addressed by a recent body of work on \emph{coalgebraic partition refinement}~\cite{Dorsch+17,Deifel+19,Wissmann+21,JacobsWissmann23}. Here, a target system is identified with a categorical construct called \emph{coalgebra} $c\colon C\to FC$ (see e.g.~\cite{Jacobs16}), where $C$ represents the state space, the \emph{functor} $F$ specifies the type of the system, and $c$ represents the dynamics. By changing the functor $F$ as a parameter, the theory accommodates many different systems such as DFAs and weighted automata.
The coalgebraic partition refinement algorithms in~\cite{Dorsch+17,Deifel+19,Wissmann+21,JacobsWissmann23} are \emph{functor-generic}: they 
apply uniformly to such a variety of systems. 

The current work is inspired by~\cite{JacobsWissmann23}
which successfully exploits Hopcroft's trick for generic coalgebraic partition refinement.
In~\cite{JacobsWissmann23}, their coalgebraic algorithm is described in parallel with its set-theoretic (or even binary-level) concrete representations, letting the latter accommodate Hopcroft's trick.
Their experimental results witnessed its superior performance, beating some existing tools that are specialised in a single type of systems. 

However, the use of Hopcroft's trick in~\cite{JacobsWissmann23} is formulated in low-level set-theoretic terms,
which seems to obscure the essence of the algorithm as well as the optimisation by Hopcroft's trick,
much like in the original paper~\cite{Hopcroft71}. Therefore, in this paper, we aim at 1) an explicit formulation of Hopcroft's trick, and 2) a categorical partition refinement algorithm that exposes an explicit data structure to which Hopcroft's trick applies.

We achieve these two goals in this paper: 1) an explicit formulation that we call \emph{Hopcroft's inequality}, and 2) a categorical algorithm that uses a \emph{fibration}. Here is an overview.

\vspace{1mm}
\myparagraph{Hopcroft's Inequality}
We identify \emph{Hopcroft's inequality} (\cref{thm:Hopcroft-ineq}) as the essence of Hopcroft's trick. 
Working on general trees with a general notion of vertex weight, it uses the classification of edges into \emph{heavy} and \emph{light} ones and bounds a sum of weights in terms of (only) the root and leaf weights.
This inequality can  be used to bound the complexity of many tree generation algorithms, including those for partition refinement.

This general theory can  accommodate different weights. We exploit this generality to systematically derive partition refinement algorithms with different complexities (\cref{sec:complexity-analysis}).

\vspace{1mm}
\myparagraph{A Fibrational Partition Refinement Algorithm}
Hopcroft's inequality does not directly apply to the existing coalgebraic partition refinement algorithms~\cite{Dorsch+17,Deifel+19,Wissmann+21,JacobsWissmann23} since the latter do not explicitly present a suitable tree structure. To address this challenge, we found the 
categorical language of \emph{fibrations}~\cite{Jacobs99} to be a convenient vehicle: it allows us to speak about the relationship between 1) an equivalence relation (an object in a fibre category) and 2) a partitioning of a state space (a mono-sink in the base category). The outcome is a partition refinement algorithm that is both \emph{abstract} (it is functor-generic and applies to a variety of systems) and \emph{concrete} (it explicitly builds a tree to which Hopcroft's inequality applies.) Our development relies on the fibrational theory of bisimilarity~\cite{HermidaJacobs98,Komorida+22}; yet ours is the first fibrational partition refinement algorithm.

More specifically, in a fibration $p\colon\cate\to\catc$,
an equivalence relation $R$ on a set $X$ is identified with an object $R\in \cate_{X}$ in the fibre over $X$
(consider the well-known fibration $\eqrel\to\sets$ of sets and equivalence relations over them).
We introduce a categorical notion of \emph{$R$-partitioning};
it allows $R\in \cate_{X}$ to induce a mono-sink (i.e.\ a family of monomorphisms) $\{ \kappa_{i} \colon C_i \monoto C\}_{i \in I}$.
The latter is identified with the set of $R$-equivalence classes. 

\cref{fig:Partitioning-intro} illustrates one iteration of our fibrational partition refinement algorithm $\PRHopcroft$ (\cref{algo:fibrational-block-specified}).
In the last step (\cref{subfig:iterRefine}),
the object $C_{01}$ is devided into three parts $C_{010}$, $C_{011}$ and $C_{012}$ along $\pull{(c \circ \kappa)}\lift{F}R$.
We call the mono-sink $\{ C_{01i} \monoto C_{01}\}_{i \in \{ 0, 1, 2\}}$ the $\pull{(c \circ \kappa)}\lift{F}R$-partitioning of $C_{01}$.
In this manner, a tree structure  explicitly emerges in the base category $\catc$.
Hopcroft's inequality directly applies to this tree, allowing us to systematically present the Hopcroft-type optimisation on the categorical level of abstraction.

We note that, at this moment, our fibrational framework (with a fibration $p\colon\cate\to\catc$) has only one example, namely the fibration $\eqrel\to\sets$ of equivalence relations over sets. 
While it is certainly desirable to have other examples, their absence does not harm the value of our fibrational framework: we do not use fibrations for additional generality (beyond functor-genericity);\footnote{In this sense, we can say that our use of fibrations is similar to some recent usages of string diagrams in \emph{specific} monoidal categories, such as in~\cite{PiedeleuKCS15,BonchiHPSZ19}.} we use them  to explicate trees in the base category (cf.\ \cref{fig:Partitioning-intro}).

\vspace{1mm}
\myparagraph{Contributions}
Summarising, our main technical contributions are as follows.
\begin{itemize}
    \item \emph{Hopcroft's inequality} that explicates the essence of Hopcroft's trick.
    \item A fibrational notion of \emph{$R$-partitioning} that turns a fibre object into a mono-sink (\cref{sec:fibrational-partitioning}).
    \item A fibrational partition refinement algorithm $\PRHopcroft$ that combines the above two (\cref{subsec:PRHopcroft}). 
    \item Functor-generic partition refinement algorithms 
        \[ \PRHopcroftEqRel{\wCard}, \quad \PRHopcroftEqRel{\wPred}, \quad \PRHopcroftEqRel{\wReach}\]
        obtained as instances of  $\PRHopcroft$ but using different weights in Hopcroft's inequality.
        The three achieve slightly different, yet comparable to the best known, complexity bounds (\cref{sec:complexity-analysis}).
\end{itemize}

\begin{figure}[t]
  \centering
  \footnotesize
  \begin{subfigure}[t]{0.28\linewidth}
      \centering
      \begin{tikzpicture}
          \node (C)    at (0, 0) {$C$};
          \node (Ca)   at (-1, 0.3) {$C_{0}$} ;
          \node (Cb)   at (-0.8, -0.3) {$C_{1}$} ;
          \node (Caa)  at (-2.7, 0.6) {$C_{00}$} ;
          \node (Cab)  at (-2.2  , -0.1) {$C_{01}$} ;
          \draw[>->] (Ca) -- (C) ;
          \draw[>->] (Cb) -- (C) ;
          \draw[>->] (Caa) -- (Ca) ;
          \draw[>->] (Cab) -- (Ca) ;
          \node (R) at (0, 1.5) {$R$} ;
          \draw[dotted] (R) -- (C) ;
      \end{tikzpicture}
      \caption{Before the iteration}
      \label{subfig:iterBefore}
  \end{subfigure}
  \begin{subfigure}[t]{0.3\linewidth}
      \centering
      \begin{tikzpicture}
          \node (C)    at (0, 0) {$C$};
          \node (Ca)   at (-1, 0.3) {$C_{0}$} ;
          \node (Cb)   at (-0.8, -0.3) {$C_{1}$} ;
          \node (Caa)  at (-2.7, 0.6) {$C_{00}$} ;
          \node (Cab)  at (-2.2  , -0.1) {$C_{01}$} ;
          \draw[>->] (Ca) -- (C) ;
          \draw[>->] (Cb) -- (C) ;
          \draw[>->] (Caa) -- (Ca) ;
          \draw[>->] (Cab) -- (Ca) ;
          \node (ckR) at (-2.2, 1.4) {$\pull{(c \circ \kappa)}\lift{F}R$} ;
          \draw[dotted] (ckR) -- (Cab) ;
          \node (cR) at (0, 1.4) {$\pull{c}\lift{F}R$} ;
          \draw[dotted] (cR) -- (C) ;
          \draw[|->] (cR) -- (-1, 1.7) -- (ckR) ;
      \end{tikzpicture}
      \caption{Refine $R$ and restrict to $C_{01}$}
      \label{subfig:iterInv}
  \end{subfigure}
  \begin{subfigure}[t]{0.38\linewidth}
      \centering
      \begin{tikzpicture}
          \node (C)    at (0, 0) {$C$};
          \node (Ca)   at (-1, 0.3) {$C_{0}$} ;
          \node (Cb)   at (-0.8, -0.3) {$C_{1}$} ;
          \node (Caa)  at (-2.7, 0.6) {$C_{00}$} ;
          \node (Cab)  at (-2.2  , -0.1) {$C_{01}$} ;
          \node[red] (Caba) at (-3.6, 0.3) {$C_{010}$} ;
          \node[red] (Cabb) at (-3.4, -0.1) {$C_{011}$} ;
          \node[red] (Cabc) at (-3.2, -0.5) {$C_{012}$} ;
          \draw[>->] (Ca) -- (C) ;
          \draw[>->] (Cb) -- (C) ;
          \draw[>->] (Caa) -- (Ca) ;
          \draw[>->] (Cab) -- (Ca) ;
          \draw[>->, red] (Caba) -- (Cab) ;
          \draw[>->, red] (Cabb) -- (Cab) ;
          \draw[>->, red] (Cabc) -- (Cab) ;
          \node (R) at (-2.2, 1.4) {$\pull{(c \circ \kappa)}\lift{F}R$} ;
          \draw[dotted] (R) -- (Cab) ;
      \end{tikzpicture}
      \caption{Refine $C_{01}$ and expand the tree}
      \label{subfig:iterRefine}
  \end{subfigure}
  \caption{An iteration in our algorithm $\PRHopcroft$ (\cref{algo:fibrational-block-specified}). \cref{subfig:iterBefore} shows an equivalence relation $R$ over $C$, and the corresponding partitioning  $C_{00}, C_{01}, C_{1}\monoto C$ of the state space $C$. (The history of refinement is recorded as a tree; this is important for complexity analysis.) In~\cref{subfig:iterInv}, the equivalence relation $R$ is refined into $\pull{c}\lift{F}R$ along the one-step transition of the system dynamics  $c$, and is further restricted to the partition $C_{01}$. In~\cref{subfig:iterRefine}, the resulting equivalence relation $\pull{(c \circ \kappa)}\lift{F}R$ over $C_{01}$ yields a partitioning of $C_{01}$, expanding the tree.
  }
  \label{fig:Partitioning-intro}
\end{figure}

\section{Hopcroft's Inequality}\label{sec:Hopcroft-trick}
We present our first contribution, \emph{Hopcroft's inequality}.
It is a novel formalisation of Hopcroft's trick in terms of rooted trees.
It also generalises the trick,  accommodating arbitrary \emph{weights} (\cref{def:weight}) besides the particular one to count the number of items in classes
that is typically and widely used (e.g.\ \cite{Hopcroft71,KanellakisSmolka90,Valmari09,JacobsWissmann23}).
\begin{mynotation}
  Let $T$ be a rooted tree. We denote
  the set of leaves by $\leaves(T)$,
  the set of vertices by $\vertex(T)$,
  the set of edges in the path from $v$ to $u$ by $\treepath(v, u)$,
  the set of children of $v \in \vertex(T)$ by $\ch(v)$, and
  the subtree whose root is $v \in \vertex(T)$ by $\subtree(v)$.
\end{mynotation}

\begin{mydefinition}[weight function] \label{def:weight}
  Let $T$ be a rooted finite tree.
  A \emph{weight function} of $T$ is a map $w \colon \vertex(T) \to \nat$ satisfying
  $\sum_{u \in \ch(v)} w(u) \le w(v)$
  for each $v \in \vertex(T)$.
  We call a weight function \emph{tight}
  if $\sum_{u \in \ch(v)} w(u) = w(v)$ for all $v \in \vertex(T) \setminus \leaves(T)$.
\end{mydefinition}

\begin{mydefinition}[heavy child choice] \label{def:heavy-child-choice}
  For a weight function $w$ of a tree $T$,
  a \emph{heavy child choice} (hcc for short) is a map $h \colon \vertex(T) \setminus \leaves(T) \to \vertex(T)$
  satisfying
  $h(v) \in \ch(v)$ and
  $w(h(v)) = \max_{u \in \ch(v)} w(u)$ for every $v \in \vertex(T) \setminus \leaves(T)$.
  We write $h(v)$ as $h_v$ and call the vertex $h_v$ a \emph{heavy child} of $v$, and a non-heavy child a \emph{light child}.
  We define $\lch_{h}(v) = \ch(v) \setminus \{ h_v \}$.
  An edge $(v, u)$ is a \emph{light edge} if $u \in \lch_h(v)$.
  We define $\lpath(v,u) = \{ e \in \treepath(v,u) \mid \text{$e$ is a light edge}\}$.
\end{mydefinition}
Note that a heavy child choice always exists but is not unique in general.

\begin{figure}[tbh]
  \footnotesize
  \centering
  \begin{tikzpicture}[scale=0.6]
      \small
      \pgfmathsetmacro{\rootx}{0}
      \pgfmathsetmacro{\rooty}{5}
      \pgfmathsetmacro{\ya}{0.75}
      \pgfmathsetmacro{\yb}{1.5}
      \pgfmathsetmacro{\yc}{2.25}
      \coordinate (root) at (\rootx, \rooty) ;
      \coordinate (n0) at (\rootx - 2, \rooty - \ya) ;
      \coordinate (n1) at (\rootx - 0, \rooty - \ya) ;
      \coordinate (n2) at (\rootx + 2, \rooty - \ya) ;
      \coordinate (n00) at (\rootx - 3, \rooty - \yb) ;
      \coordinate (n01) at (\rootx - 2, \rooty - \yb) ;
      \coordinate (n10) at (\rootx - 1, \rooty - \yb) ;
      \coordinate (n11) at (\rootx - 0, \rooty - \yb) ;
      \coordinate (n12) at (\rootx + 1, \rooty - \yb) ;
      \coordinate (n20) at (\rootx + 2, \rooty - \yb) ;
      \coordinate (n200) at (\rootx + 1.5, \rooty - \yc) ;
      \coordinate (n201) at (\rootx + 2.5, \rooty - \yc) ;
      \draw[heavy] (root) -- (n0) ;
      \draw[light] (root) -- (n1) ;
      \draw[light] (root) -- (n2) ;
      \draw[light] (n0) -- (n00) ;
      \draw[heavy] (n0) -- (n01) ;
      \draw[heavy] (n1) -- (n10) ;
      \draw[light] (n1) -- (n11) ;
      \draw[light] (n1) -- (n12) ;
      \draw[heavy] (n2) -- (n20) ;
      \draw[light] (n20) -- (n200) ;
      \draw[heavy] (n20) -- (n201) ;
      \draw[fill=white] (root) circle[radius=3mm];
      \node[circle] at  (root) {$36$} ;
      \draw[fill=white] (n0) circle[radius=3mm];
      \node[circle] at  (n0) {$14$} ;
      \draw[fill=white] (n1) circle[radius=3mm];
      \node[circle] at  (n1) {$14$} ;
      \draw[fill=white] (n2) circle[radius=3mm];
      \node[circle] at  (n2) {$7$} ;
      \draw[fill=white] (n00) circle[radius=3mm];
      \node[circle] at  (n00) {$5$} ;
      \draw[fill=white] (n01) circle[radius=3mm];
      \node[circle] at  (n01) {$7$} ;
      \draw[fill=white] (n10) circle[radius=3mm];
      \node[circle] at  (n10) {$5$} ;
      \draw[fill=white] (n11) circle[radius=3mm];
      \node[circle] at  (n11) {$2$} ;
      \draw[fill=white] (n12) circle[radius=3mm];
      \node[circle] at  (n12) {$3$} ;
      \draw[fill=white] (n20) circle[radius=3mm];
      \node[circle] at  (n20) {$7$} ;
      \draw[fill=white] (n200) circle[radius=3mm];
      \node[circle] at  (n200) {$2$} ;
      \draw[fill=white] (n201) circle[radius=3mm];
      \node[circle] at  (n201) {$4$} ;
  \end{tikzpicture}
  \begin{tikzpicture}[scale=0.6]
      \small
      \pgfmathsetmacro{\rootx}{0}
      \pgfmathsetmacro{\rooty}{5}
      \pgfmathsetmacro{\ya}{0.75}
      \pgfmathsetmacro{\yb}{1.5}
      \pgfmathsetmacro{\yc}{2.25}
      \coordinate (root) at (\rootx, \rooty) ;
      \coordinate (n0) at (\rootx - 2, \rooty - \ya) ;
      \coordinate (n1) at (\rootx - 0, \rooty - \ya) ;
      \coordinate (n2) at (\rootx + 2, \rooty - \ya) ;
      \coordinate (n00) at (\rootx - 3, \rooty - \yb) ;
      \coordinate (n01) at (\rootx - 2, \rooty - \yb) ;
      \coordinate (n10) at (\rootx - 1, \rooty - \yb) ;
      \coordinate (n11) at (\rootx - 0, \rooty - \yb) ;
      \coordinate (n12) at (\rootx + 1, \rooty - \yb) ;
      \coordinate (n20) at (\rootx + 2, \rooty - \yb) ;
      \coordinate (n200) at (\rootx + 1.5, \rooty - \yc) ;
      \coordinate (n201) at (\rootx + 2.5, \rooty - \yc) ;
      \draw[heavy] (root) -- (n0) ;
      \draw[light] (root) -- (n1) ;
      \draw[light] (root) -- (n2) ;
      \draw[light] (n0) -- (n00) ;
      \draw[heavy] (n0) -- (n01) ;
      \draw[heavy] (n1) -- (n10) ;
      \draw[light] (n1) -- (n11) ;
      \draw[light] (n1) -- (n12) ;
      \draw[heavy] (n2) -- (n20) ;
      \draw[light] (n20) -- (n200) ;
      \draw[heavy] (n20) -- (n201) ;
      \draw[fill=white] (root) circle[radius=3mm];
      \node[circle] at  (root) {$36$} ;
      \draw[fill=white] (n0) circle[radius=3mm];
      \node[circle] at  (n0) {$15$} ;
      \draw[fill=white] (n1) circle[radius=3mm];
      \node[circle] at  (n1) {$14$} ;
      \draw[fill=white] (n2) circle[radius=3mm];
      \node[circle] at  (n2) {$7$} ;
      \draw[fill=white] (n00) circle[radius=3mm];
      \node[circle] at  (n00) {$5$} ;
      \draw[fill=white] (n01) circle[radius=3mm];
      \node[circle] at  (n01) {$10$} ;
      \draw[fill=white] (n10) circle[radius=3mm];
      \node[circle] at  (n10) {$9$} ;
      \draw[fill=white] (n11) circle[radius=3mm];
      \node[circle] at  (n11) {$2$} ;
      \draw[fill=white] (n12) circle[radius=3mm];
      \node[circle] at  (n12) {$3$} ;
      \draw[fill=white] (n20) circle[radius=3mm];
      \node[circle] at  (n20) {$7$} ;
      \draw[fill=white] (n200) circle[radius=3mm];
      \node[circle] at  (n200) {$2$} ;
      \draw[fill=white] (n201) circle[radius=3mm];
      \node[circle] at  (n201) {$5$} ;
  \end{tikzpicture}
  \caption{
    Examples of rooted trees, each with a weight function and an hcc.
    The heavy children are indicated by thick edges.
    Thin edges are light edges.
  }
  \label{fig:weighted-heavy-chosen-tree}
\end{figure}
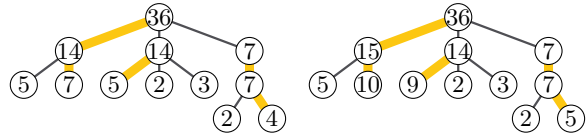

Examples are in \cref{fig:weighted-heavy-chosen-tree}; the weight on the left is not tight while the right one is tight.

In the rest of this section, our technical development is towards  Hopcroft's inequality in \cref{thm:Hopcroft-ineq}.
It gives an upper bound for a sum of weights---we only count those for light children,
which is the core of the optimisation in Hopcroft's partition refinement algorithm~\cite{Hopcroft71}---in terms of weights of the root and the leaves.
This upper bound makes no reference to the tree's height or internal weights,
making it useful for complexity analysis of tree generation algorithms.

The following lemma crucially relies on the definition of weight function.

\begin{mylemmarep}\label{lem:path-tate-yoko-general}
  Let $T$ be a finite tree with a root $r$, $w$ be a weight function of $T$,
  and $S$ be an arbitrary set of edges of $T$.
  Then
  $\sum_{v \in \vertex(T)} \sum_{\substack{u \in \ch(v) \\ (v, u) \not\in S}} w(u) \ge \sum_{l \in \leaves(T)}  \bigl| \treepath(r, l) \setminus S \bigr| \cdot w(l)$
  holds.
  The equality holds when $w$ is tight.
\end{mylemmarep}
\begin{proof}
  Let $r$ be the root of $T$.
  We prove by the induction on the size of $T$.
  
  (\textbf{Base case}).
  $T$ is a tree whose vertex is only $r$.
  Thus we have
  \begin{equation*}
      \sum_{v \in \vertex(T)} \sum_{\substack{u \in \ch(v) \\ (v, u) \not\in S}} w(u)
      = \sum_{\substack{u \in \varnothing = \ch(r) \\ (v, u) \not\in S}} w(u)
      = 0
      = \bigl| \varnothing \bigr| \cdot w(r)
      = \sum_{l \in \leaves(T)}  \bigl| \treepath(r, l) \setminus S \bigr| \cdot w(l).
  \end{equation*}

  (\textbf{Induction step}).
  Let $U = \ch(r) \cap S$ and $W = \ch(r) \setminus S$.
  The following calculation shows the inequality:
  \begin{align*}
      & \sum_{v \in \vertex(T)} \sum_{\substack{u \in \ch(v) \\ (v, u) \not\in S}} w(u)
      \\
      & = \sum_{r' \in W} w(r')
        + \sum_{r' \in U \cup W} \sum_{v \in \vertex(\subtree(r'))} \sum_{\substack{u \in \ch(v) \\ (v, u) \not\in S}} w(u)
      \\
      & \ge \sum_{r' \in W} w(r')
        + \sum_{r' \in U \cup W} \sum_{l \in \leaves(\subtree(r'))}  \bigl| \treepath(r', l) \setminus S \bigr| \cdot w(l)
      \\
      & \hspace{7.6cm} \text{by the induction hypothesis}
      \\
      & = \sum_{r' \in U} \sum_{l \in \leaves(\subtree(r'))}  \bigl| \treepath(r', l) \setminus S \bigr| \cdot w(l) \\
      & \hspace{2em} + \sum_{r' \in W} \left( w(r') + \sum_{l \in \leaves(\subtree(r'))}  \bigl| \treepath(r', l) \setminus S \bigr| \cdot w(l) \right)
      \\
      & \ge \sum_{r' \in U} \sum_{l \in \leaves(\subtree(r'))}  \bigl| \treepath(r', l) \setminus S \bigr| \cdot w(l) \\
      & \hspace{2em} + \sum_{r' \in W} \left( \sum_{l \in \leaves(\subtree(r'))} w(l) + \sum_{l \in \leaves(\subtree(r'))}  \bigl| \treepath(r', l) \setminus S \bigr| \cdot w(l) \right)
      \\
      & \hspace{7.6cm} \text{by $\displaystyle w(r') \ge \sum_{l \in \leaves(\subtree(r'))} w(l)$}
      \\
      & = \sum_{r' \in U} \sum_{l \in \leaves(\subtree(r'))}  \bigl| \treepath(r', l) \setminus S \bigr| \cdot w(l) \\
      & \hspace{2em} + \sum_{r' \in W} \sum_{l \in \leaves(\subtree(r'))}  \left( w(l) +  \bigl| \treepath(r', l) \setminus S \bigr| \cdot w(l) \right)
      \\
      & = \sum_{r' \in U} \sum_{l \in \leaves(\subtree(r'))}  \bigl| \treepath(r, l) \setminus S \bigr| \cdot w(l)
          + \sum_{r' \in W} \sum_{l \in \leaves(\subtree(r'))}  \bigl| \treepath(r, l) \setminus S \bigr| \cdot w(l)
      \\
      & = \sum_{l \in \leaves(T)}  \bigl| \treepath(r, l) \setminus S \bigr| \cdot w(l).
  \end{align*}

  When $w$ is tight,
  we have $w(r') = \sum_{l \in \leaves(\subtree(r'))} w(l)$ for every $r' \in W$,
  and the equality holds.
\end{proof}

\cref{lem:path-tate-yoko} is our first key lemma; we use \cref{lem:path-tate-yoko-general} in its proof.
It relates the sum of weights of the light children---for which we aim to give an upper bound in \cref{thm:Hopcroft-ineq}---with the leaf weights and (roughly) the tree height.

\begin{mylemma}\label{lem:path-tate-yoko}
  Let $T$ be a finite tree with a root $r$, $w$ be a weight function of $T$, and $h$ be an hcc for $w$.
  Then the following inequality holds. The equality holds when $w$ is tight.
  \begin{equation}\label{eq:tate-yoko}\textstyle
      \sum_{v \in \vertex(T)} \sum_{u \in \lch_h(v)} w(u)
      \;\ge\; \sum_{l \in \leaves(T)} 
      \bigl| \lpath(r, l) \bigr| \cdot w(l).
  \end{equation}
\end{mylemma}

For the right tree in \cref{fig:weighted-heavy-chosen-tree},
the left-hand side of (\ref{eq:tate-yoko}) is $(14 + 7) + 5 + (2 + 3) + 0 + 2 = 33$, and
the right-hand side is $1 \times 5 + 0 \times 10 + 1 \times 9 + 2 \times 2 + 2 \times 3 + 2 \times 2 + 1 \times 5 = 33$.

The inequality in~(\ref{eq:tate-yoko}) is the opposite of what we want (namely an upper bound for the left-hand side).
We thus force an equality using
\emph{tightening}.
\begin{mydefinition}[tightening]
  Let  $w$ be  a weight function of 
  a rooted finite tree $T$, and  $h$ be its heavy child choice.
  The \emph{tightening}  $w' \colon V(T) \to \nat$  of $w$ along $h$ is defined recursively by
  \begin{equation*}
      w'(u) = \left\{
          \begin{array}{ll}
              w(u) & \text{if $u$ is the root of $T$}
              \\
              w'(v) - \sum_{u' \in \lch_h(v)} w(u') & \text{if $u = h_v$ for the parent $v$ of $u$}
              \\
              w(u) & \text{otherwise.}
          \end{array}
      \right.
  \end{equation*}
\end{mydefinition}
In \cref{fig:weighted-heavy-chosen-tree}, the weight function of the right tree
is a tightening of that of the left tree. We observe that tightening maintains a heavy child choice:

\begin{mylemma}\label{lem:property-strictification}
  Let $T$ be a rooted finite tree, $w$ be a weight function of $T$, $h$ be an hcc for $w$,
  and $w'$ be the tightening of $w$ along $h$.
  The following hold.
  \begin{enumerate}
      \item The map $w'$ is a tight weight function of $T$.
      \item The map $h$ is also a heavy child choice for $w'$.
      \item For the root $r$, $w(r) = w'(r)$ holds, and for each $v \in \vertex(T)$, $w(v) \le w'(v)$ holds.
  \end{enumerate}
\end{mylemma}

Our second key lemma towards \cref{thm:Hopcroft-ineq} is as follows,
bounding $|\lpath(r, v)|$ that occurs on the right in~(\ref{eq:tate-yoko}).
Its proof is by what is commonly called \emph{Hopcroft's trick}~\cite{Hopcroft71,BerkholzBG17}:
it observes that, along a light edge, weights decay at least by $1/2$.
\begin{mylemmarep}\label{lem:light-edge-bound}
  Let $T$ be a finite tree with a root $r$,
  $w$ be a weight of $T$, and
  $h$ be an hcc for $w$.
  For each vertex $v \in \vertex(T)$ with $w(v) \ne 0$,
  the following inequality holds: $|\lpath(r, v)| \le \log_2 w(r) - \log_2 w(v)$.
\end{mylemmarep}
\begin{proof}
  Let $\langle (v_1, u_1), \dots , (v_m, u_m) \rangle$ be
  the sequence of the edges of $\lpath(r, v)$ in the order from $r$ to $v$.
  Since $u_i \ne h_{v_i}$,
  we have $2 \cdot w(u_i) \le w(v_i)$.
  We also have $w(v_{i+1}) \le w(u_i)$.
  Hence, the following inequality holds:
  \begin{align*}
      w(r) & \ge w(v_1) \ge 2 \cdot w(u_1)
      \\
      & \ge 2 \cdot w(v_2) \ge 2^2 \cdot w(u_2)
      \\
      & \dots
      \\
      & \ge 2^{m-1} \cdot w(v_m) \ge 2^{m} \cdot w(u_m)
      \\
      & \ge 2^{m} \cdot w(v).
  \end{align*}
  Taking the logarithm of both sides yields $|\lpath(r,v)| = m \le \log_2 w(r) - \log_2 w(v)$.
\end{proof}

We combine \cref{lem:light-edge-bound} and \cref{lem:path-tate-yoko} (its equality version; we can use it via tightening) to obtain Hopcroft's inequality.
It bounds a sum of weights by the root and leaf weights.
\begin{mytheoremrep}[Hopcroft's inequality]\label{thm:Hopcroft-ineq}
    Let $T$ be a finite tree with root $r$,
    $w$ be a weight function of $T$,
    and $h$ be a heavy child choice for $w$.
    The following inequality holds.
    \begin{equation}\label{eq:Hopcroft-ineq}
        \sum_{v \in \vertex(T)} \sum_{u \in \lch_h(v)} w(u)
        \le
        w(r) \log_2 w(r) - \sum_{\substack{ l \in \leaves(T) \\ w(l) \ne 0 }} w(l) \log_2 w(l)
    \end{equation}
\end{mytheoremrep}
\begin{proof}
    Let $w'$ be the tightening of $w$.
    By \cref{lem:property-strictification},
    $w'$ is a tight weight function, and
    $h$ is also a heavy child choice for $w'$.
    If we have
    \begin{equation}\label{eq:Hopcroft-inequality-strict}
        \sum_{v \in \vertex(T)} \sum_{u \in \lch_h(v)} w'(u)
        \;\le\;
        w'(r) \log_2 w'(r) - \sum_{ l \in \leaves(T), w'(l) \ne 0 } w'(l) \log_2 w'(l)
    \end{equation}
    then the desired inequality holds:
    \begin{align*}
        \sum_{v \in \vertex(T)} \sum_{u \in \lch_h(v)} w(u)
        & \le \sum_{v \in \vertex(T)} \sum_{u \in \lch_h(v)} w'(u)
        & \text{\cref{lem:property-strictification}}
        \\
        & \le w'(r) \log_2 w'(r) - \sum_{\substack{l \in \leaves(T) \\ w'(l) \ne 0}} w'(l) \log_2 w'(l)
        \\
        & \le w(r) \log_2 w(r) - \sum_{\substack{l \in \leaves(T) \\ w(l) \ne 0}} w(l) \log_2 w(l).
        & \text{\cref{lem:property-strictification}}
    \end{align*}

    Now, our goal is to show (\ref{eq:Hopcroft-inequality-strict}).
    It is proven by the following calculation:
    \begin{align*}
        & \sum_{v \in \vertex(T)} \sum_{u \in \lch_h(v)} w'(u)
        \\
        & = \sum_{l \in \leaves(T)} \sum_{(v, u) \in \lpath(r, l)} w'(l)
        & \text{$w'$: tight, and \cref{lem:path-tate-yoko}}
        \\
        & = \sum_{\substack{l \in \leaves(T) \\ w'(l) \ne 0}} \sum_{(v, u) \in \lpath(r, l)} w'(l)
        &
        \\
        & \le \sum_{\substack{l \in \leaves(T) \\ w'(l) \ne 0}} (\log_2 w'(r) - \log_2 w'(l)) \cdot w'(l)
        & \text{\cref{lem:light-edge-bound}}
        \\
        & = \left( \sum_{\substack{l \in \leaves(T) \\ w'(l) \ne 0}} w'(l) \right) \log_2 w'(r) - \sum_{\substack{l \in \leaves(T) \\ w'(l) \ne 0}} w'(l) \log_2 w'(l)
        \\
        & = w'(r) \log_2 w'(r) - \sum_{\substack{l \in \leaves(T) \\ w'(l) \ne 0}} w'(l) \log_2 w'(l).
    \end{align*}
\end{proof}

For complexity analysis, we use Hopcroft's inequality in the following form.
Assume that a tree generation algorithm takes $t(v)$ time to generate all the children (both heavy and light) of $v$.
If there exists $K$ such that $t(v)$ is bounded by $K$ times the sum of all light children, then the time to generate the whole tree is bounded by $K w(r) \log_2 w(r)$.
\begin{mycorollary}\label{cor:time-estimation}
    Let $T$ be a rooted finite tree with root $r$,
    $w$ be a weight function of $T$,
    and $h$ be a heavy child choice for $w$.
    If a map $t \colon \vertex(T) \to \nat$ satisfies that
    there exists a constant $K \in \nat$ such that
    $t(v) \le K \sum_{u \in \lch_h(v)} w(u)$
    for every $v \in \vertex(T)$,
    then the sum of $t(v)$ is bounded by $K w(r) \log_2 w(r)$, that is
    $\sum_{v \in \vertex(T)} t(v) \le K w(r) \log_2 w(r)$.
\end{mycorollary}

\begin{myremark}
 Further adaptations of Hopcroft's trick are pursued in the literature, e.g.\ in~\cite{ValmariFranceschinis10}, where the notion of heavy child choice is relaxed with an extra parameter $\alpha \in [1/2, 1)$. Our theory can easily be extended to accommodate $\alpha$, in which case the above description corresponds to the special case 
 with $\alpha=1/2$. Details are deferred to another venue.
\end{myremark}

\section{Categorical Preliminaries}\label{sec:preliminaries}
The rest of the paper is about our second contribution, namely a functor-generic partition refinement (PR) algorithm  optimised by an explicit use of Hopcroft's inequality (\cref{thm:Hopcroft-ineq}). It is given by our novel formulation of coalgebraic PR algorithms in fibrational terms. Here we shall review some necessary categorical preliminaries.

We use categorical formalisation of intersections and unions. 

\begin{mydefinition}
  For monomorphisms $m \colon A \monoto C$ and $n \colon B \monoto C$ in $\catc$,
  the \emph{intersection} $m \cap n \colon A \cap B \monoto C$
  and the \emph{union} $m \cup n \colon A \cup B \monoto C$
  are defined by the pullback and pushout, respectively:
  \begin{equation*}
      \begin{tikzcd}[row sep=small, column sep=small]
          A \cap B
          \ar[d, >->, "\pi_1"']
          \ar[r, >->, "\pi_2"]
          \ar[rd, phantom, "\usebox\pullback", very near start]
          &
          B
          \ar[d, >->, "n"]
          \\
          A
          \ar[r, >->, "m"]
          &
          C 
      \end{tikzcd}
      \quad \text{and} \quad
      \begin{tikzcd}[row sep=small, column sep=small]
          A \cap B
          \ar[d, >->, "\pi_1"']
          \ar[r, >->, "\pi_2"]
          \ar[rd, phantom, "\usebox\pushout", very near end]
          &
          B
          \ar[d, >->]
          \\
          A
          \ar[r, >->]
          &
          A \cup B 
      \end{tikzcd}.
  \end{equation*}
\end{mydefinition}
We say $m \colon A \monoto C$ and $m' \colon A' \monoto C$ are \emph{equivalent}
if there is an isomorphism $\phi \colon A \to A'$ such that $m = m' \circ \phi$.
The set $\sub(\catc)_{C}$ of equivalence classes of monomorphisms whose codomains are $C$ forms a lattice, assuming enough limits and colimits.

\subsection{Fibrations}\label{subsec:prelimFib}

A fibration $p \colon \cate \to \catc$ is a functor satisfying some axioms.
When $p(R)=C$ for an object $R \in \cate$ and an object $C \in \catc$,
we see that $R$ equips $C$ with some information,
e.g.\ a predicate, a relation, a topology, etc.
The main example in this paper is the fibration $\eqrel\to\sets$ where $C$ is a set and $R$ is an equivalence relation over $C$. 

Fibrational constructs that are the most relevant to us are the \emph{inverse image} $\pull{f}(R')$ and the \emph{direct image} $\push{f}(R)$ along a morphism $f\colon S\to S'$ in $\catc$. In the case of $\eqrel\to\sets$, these are computed as follows.
\begin{displaymath}
    \begin{minipage}[b]{0.49\linewidth}
        \centering
        \begin{tikzpicture}
            \pgfmathsetmacro{\x}{3}
            \node (S)  at (0,0) {$S$} ;
            \node (S') at (\x,0) {$S'$} ;
            \draw[->] (S) to node[above] {$f$} (S') ;
            \node (R') at (\x, 1) {$R'$} ;
            \node (pullfR') at (0, 1) {\begin{minipage}{3.5cm}$\pull{f}(R') = $ \\ $\{ (x,y) \mid (fx, fy) \in R'\}$\end{minipage}} ;
            \draw[|->] (R') to node[above]{$\pull{f}$} (pullfR') ;
            \draw[dotted, gray] (R') to (S') ;
            \draw[dotted, gray] (pullfR') to (S) ;
        \end{tikzpicture}
    \end{minipage}
    \quad
    \begin{minipage}[b]{0.49\linewidth}
        \centering
        \begin{tikzpicture}
            \pgfmathsetmacro{\x}{3}
            \node (S)  at (0,0) {$S$} ;
            \node (S') at (\x,0) {$S'$} ;
            \draw[->] (S) to node[above] {$f$} (S') ;
            \node (R) at (0, 1) {$R$} ;
            \node (pushfR) at (\x, 1) {\begin{minipage}{3.5cm}$\push{f}(R) = $ \\ $\{ (fx,fy) \mid (x, y) \in R\}$\end{minipage}} ;
            \draw[|->] (R) to node[above] {$\push{f}$} (pushfR) ;
            \draw[dotted, gray] (R) to (S) ;
            \draw[dotted, gray] (pushfR) to (S') ;
        \end{tikzpicture}
    \end{minipage}
\end{displaymath}

In what follows we introduce some basics of fibrations; they formalise the intuition above. 
For details, see e.g.\ \cite{Jacobs99}.
\begin{mydefinition}[fibration]
  Let $p \colon \cate \to \catc$ be a functor.
  A morphism $f \colon P \to R$ in $\cate$ is \emph{Cartesian} if
  for any $g \colon Q \to R$ in $\cate$ with $pg = pf \circ v$ for some $v \colon pQ \to pP$,
  there exists a unique $h \colon Q \to P$ in $\cate$ above $v$ (i.e.\ $ph = v$) with $f \circ h = g$.
  The functor $p$ is a \emph{fibration} if
  for each $R \in \cate$ and $u \colon C \to pR$ in $\catc$,
  there are an object $\pull{u}R$ and a Cartesian morphism $\dot{u}(R) \colon \pull{u}R \to R$ in $\cate$. See below.
  \begin{equation*}
    \begin{tikzpicture}[scale=0.8]
        \node (E) at (0,2) {$\cate$} ;
        \node (C) at (0,0) {$\catc$} ;
        \draw[->] (E) -- node[left] {$p$} (C) ;
        \node (Q)  at (1, 2.5) {$Q$} ;
        \node (uR) at (3, 1.6) {$\pull{u}R$} ;
        \node (R)  at (5, 1.6) {$R$} ;
        \draw[->] (Q) to[out=0, in=140] node[above] {$g$} (R) ;
        \draw[->] (uR) -- node[below] {$\dot{u}R$} (R) ;
        \draw[->,dashed] (Q) -- node[below left] {$h$} (uR) ;
        \node (B) at (1,  0.5) {$B$} ;
        \node (C) at (3, -0.4) {$C$} ;
        \node (D) at (5, -0.4) {$D$} ;
        \draw[->] (B) to[out=0, in=140] node[above] {$u \circ v$} (D) ;
        \draw[->] (C) -- node[below] {$u$} (D) ;
        \draw[->] (B) -- node[below left] {$v$} (C) ;
    \end{tikzpicture}
  \end{equation*}
\end{mydefinition}

The category $\cate$ is called the \emph{total category}
and the category $\catc$ is called the \emph{base category} of the fibration.
For an object $C \in \catc$, the objects in $\cate$ above $C$ form a category $\cate_C$,
called the \emph{fibre category} above $C$.
The fibre category $\cate_C$ is the category of ``equivalence relations'' on $C$.
\begin{mydefinition}[fibre category]
  Let $p \colon \cate \to \catc$ be a fibration and $C \in \catc$.
  The \emph{fibre category} $\cate_C$ over $C$ is the subcategory of $\cate$
  whose objects are defined by $\obj(\cate_C) = \{ R \in \cate \mid pR = C \}$,
  and morphisms are defined by $\cate_C(Q,R) = \{ f \in \cate(Q,R) \mid pf = \idmorph_C \}$
  for $Q, R \in \obj(\cate_C)$.
\end{mydefinition}

\begin{myexample}[$\eqrel\to\sets$ is a fibration]\label{ex:eqrelIsAFib}
  Let $\eqrel$ be the category of equivalence relations.
  The objects of $\eqrel$ are pairs $(S, R)$ of a set $S$ and an equivalence relation $R$ on $S$.
  A morphism $f \colon (S, R) \to (S', R')$ in $\eqrel$ is a function $f \colon S \to S'$
  satisfying $(f(x), f(y)) \in R'$ for all $(x, y) \in R$.
  We sometimes write just $R$ for $(S, R)$ when no confusion arises.
  The functor $p \colon \eqrel \to \sets$ defined by $p(S, R) = S$ is a fibration.
\end{myexample}

For a morphism $u \colon C \to D$ in the base category of a fibration,
the map $\pull{u} \colon \obj(\cate_D) \to \obj(\cate_C)$ extends to a functor
$\pull{u} \colon \cate_D \to \cate_C$ between fibre categories.
We call the functor $\pull{u} \colon \cate_D \to \cate_C$ an \emph{inverse image} functor.

Given a fibration $\cate \to \catc$
and an endofunctor $F \colon \catc \to \catc$ on the base category,
if $R \in \cate$ is above $C \in \catc$,
we would like to get an object in $\cate$ above $FC$.
A lifting of $F$ specifies the choice of an object above $FC$.
\begin{mydefinition}[lifting, fibred lifting]\label{def:lifting}
  Let $p \colon \cate \to \catc$ be a fibration and $F \colon \catc \to \catc$ be a functor.
  A \emph{lifting} of $F$ is a functor $\lift{F} \colon \cate \to \cate$ with $p \circ \lift{F} = F \circ p$.
  A functor $\lift{F} \colon \cate \to \cate$ is a \emph{fibred lifting} of $F$ if $\lift{F}$ is a lifting of $F$ and preserves Cartesian morphisms.
\end{mydefinition}

When $\lift{F} \colon \cate \to \cate$ is a fibred lifting of $F \colon \catc \to \catc$, for $f \colon C \to D$ in $\catc$ and $R \in \cate_D$,
we have $\lift{F}(\pull{f} R) = \pull{(Ff)}(\lift{F}R)$ in $\cate_{FC}$.
An important example of a lifting is a relation lifting.
\begin{mydefinition}[relation lifting \cite{Jacobs16}]\label{def:relation-lifting}
  Let $F \colon \sets \to \sets$ be a weak pullback preserving functor.
  We define a lifting $\rellift{F} \colon \eqrel \to \eqrel$ of $F$
  along the fibration $p \colon \eqrel \to \sets$,
  called the \emph{relation lifting} of $F$, as follows.
  For an object $(C, R) \in \eqrel$,
  there is the inclusion $\langle r_1, r_2 \rangle \colon R \monoto C \times C$.
  We define the relation lifting on the object $R$ by $\rellift{F}(R) = \image \langle F r_1, F r_2 \rangle$,
  where $\image \langle F r_1, F r_2 \rangle$ is the image factorisation.
  By the assumption that $F$ preserves weak pullbacks, we can show that $\rellift{F}(R)$ is an equivalence relation.
  $\rellift{F}$ can be extended to a functor.
  \begin{equation*}
    \begin{tikzpicture}[scale=0.8]
        \node (FR)    at (0,0) {$FR$};
        \node (FCFC)  at (4,0) {$FC \times FC$};
        \node (Im)    at (2,-1) {$\image \langle F r_1, F r_2 \rangle$};
        \node (RelFR) at (2,-1.5) {$=\rellift{F}(R)$} ;
        \draw[->] (FR) -- node[above] {$\langle F r_1, F r_2 \rangle$} (FCFC);
        \draw[->>] (FR) -- (Im);
        \draw[>->] (Im) -- (FCFC);
    \end{tikzpicture}
  \end{equation*}
\end{mydefinition}

In this paper, we deal with a restricted class of fibrations, called $\clat$-fibrations.
\begin{mydefinition}[$\clat$-fibration]
  A fibration $p \colon \cate \to \catc$ is a \emph{$\clat$-fibration} if each fibre  $\cate_C$ is a complete lattice and
  each inverse image functor $\pull{u} \colon \cate_D \to \cate_{C}$ preserves meets $\sqcap$.
\end{mydefinition}
For a $\clat$-fibration, there always exists the left adjoint $\push{u} \colon \cate_{C} \to \cate_{D}$
to an inverse image functor $\pull{u}$, as is well-known (cf.\ Freyd's adjoint functor theorem). 
The functor $\push{u}$ is defined by
$\push{u}(P) = \bigsqcap \{ R \in \cate_{D} \mid P \sqsubseteq \pull{u}(R) \}$ on objects.
We call $\push{u}$ a \emph{direct image} functor.

\begin{myexample}[$\eqrel\to\sets$ is a $\clat$-fibration]\label{ex:eqrelIsACLatFib}
  The functor $p \colon \eqrel \to \sets$ from \cref{ex:eqrelIsAFib} is a $\clat$-fibration.
  We describe the inverse image functor $\pull{f}$
  and the direct image functor $\push{f}$ for a function $f \colon S \to S'$.
  For an equivalence relation $R'$ on $S'$,
  the inverse image $\pull{f}(R')$
  is the equivalence relation $\{ (x,y) \in S \times S \mid (f(x), f(y)) \in R' \}$ on $S$.
  For an equivalence relation $R$ on $S$,
  the direct image $\push{f}(R)$
  is the \emph{equivalence closure} of the relation
  $\bigl\{ (f(x),f(y)) \in S' \times S' \,\big|\, (x, y) \in R \bigr\}$.
\end{myexample}

\subsection{Coalgebras and Bisimulations}
Coalgebras are widely used as a generalisation of state-based systems \cite{Jacobs16,Rutten00}.

\begin{mydefinition}[$F$-coalgebra]
    Let $\catc$ be a category and $F \colon \catc \to \catc$ be an endofunctor.
    An \emph{$F$-coalgebra} is a pair $(C, c)$ of an object $C \in \catc$ and a morphism $c \colon C \to FC$.
\end{mydefinition}

For an $F$-coalgebra $c \colon C \to FC$,
$F$ specifies the type of the system,
the carrier object $C$ represents the ``set of states'' of the system,
and $c$ represents the transitions in the system.
When $\catc = \sets$, for an $F$-coalgebra $c \colon C \to FC$ and a state $x \in C$,
the element $c(x) \in FC$ represents properties (e.g.\ acceptance) and successors of $x$.

A major benefit of coalgebras is that their theory is \emph{functor-generic}: by changing a functor $F$, the same theory uniformly applies to a vast variety of systems.
\begin{myexample}
    \label{example:F-coalg}
    We describe some $F$-coalgebras for functors $F$ on $\sets$.
    \begin{enumerate}
        \item For the powerset functor $\powset$,
            a $\powset$-coalgebra $c \colon C \to \powset C$ is a \emph{Kripke frame}.
            For a state $x \in C$, $c(x) \in \powset C$ is the set of successors of $x$.
        \item Let $\Sigma$ be an alphabet and
             $\nondet_{\Sigma} = 2 \times (\powset \blank)^{\Sigma}$.
            An $\nondet_{\Sigma}$-coalgebra $c \colon C \to \nondet_{\Sigma}C$ is a \emph{non-deterministic automaton} (NA).
            For a state $x \in C$, let $(b, t) = c(x) \in 2 \times (\powset C)^{\Sigma}$.
            The state $x$ is accepting iff $b = 1$,
            and there is a transition $x \xrightarrow{a} y$ in the NA
            iff $y \in t(a)$.
        \item The distribution functor $\distr$ is defined on a set $X$ to be
            $\distr X = \{ d \colon X \to [0,1] \mid
            \{ x \in X \mid f(x) \ne 0\} \text{ is finite and } \sum_{x \in X} d(x) = 1 \}$.
            A $\distr$-coalgebra $c \colon C \to \distr C$ is a \emph{Markov chain}.
            For a state $x$, $c(x) \in \distr C$ is a probability distribution $C \to [0,1]$,
            which represents the probabilities of transitions to successor states of $x$.
    \end{enumerate}
\end{myexample}

We are interested in how similar two states of a state-transition system are. 
We consider two states to be similar if one state can mimic the transitions of the other.
\emph{Bisimilarity} by Park~\cite{Park81} and Milner~\cite{Milner89}
is a notion that captures such behaviour of states.
Hermida and Jacobs~\cite{HermidaJacobs98} formulated bisimilarity as a coinductive relation on a coalgebra, using a fibration.

\begin{mydefinition}[bisimulations and the bisimilarity]\label{def:fibrational-bisim}
    Let $p \colon \cate \to \catc$ be a $\clat$-fibration, 
    $F \colon \catc \to \catc$ be a functor,
    $c \colon C \rightarrow FC$ be an $F$-coalgebra 
    and $\lift{F}$ be a lifting of $F$.
    An \emph{$(F,\lift{F})$-bisimulation} is a $\pull{c} \circ \lift{F}$-coalgebra in $\cate_C$,
    that is an object $R \in \cate_C$ with $R \sqsubseteq \pull{c}(\lift{F} (R))$ since a morphism in $\cate_{C}$ is a relation $(\sqsubseteq)$.
    By the Knaster--Tarski theorem,
    there exists the greatest $(F, \lift{F})$-bisimulation $\nu(\pull{c} \circ \lift{F})$
    with respect to the order of $\cate_C$,
    and it is called the \emph{$(F,\lift{F})$-bisimilarity}.
\end{mydefinition}

In the above definition, the choice of $\lift{F}$ determines a notion of bisimulation.
The relation lifting $\rellift{F}$ (\cref{def:relation-lifting}) is often used as a lifting of $F$.
For all the functors we consider, the bisimilarity wrt.\ $\rellift{F}$ coincides with the \emph{behavioural equivalence}, another well-known notion of bisimilarity~\cite[\S{}4.5]{Jacobs16}.
\begin{myexample}
    We illustrate $(F, \rellift{F})$-bisimilarities
    (also called \emph{logical $F$-bisimilarity}~\cite{Jacobs16})
    for  $F$  in \cref{example:F-coalg}.
    Let $C \in \sets$ and $R \in \eqrel_C$.
    \begin{enumerate}
        \item ($F = \powset$).
            The $(\powset, \rellift{\powset})$-bisimilarity $\nu(\pull{c} \circ \rellift{\powset})$
            for a $\powset$-coalgebra $c \colon C \to \powset C$
            is the maximum relation $R$ on $C$ such that if $(x, y) \in R$ then
            \begin{itemize}
                \item for every $x' \in c(x)$, there is $y' \in c(y)$ such that $(x', y') \in R$, and
                \item for every $y' \in c(y)$, there is $x' \in c(x)$ such that $(x', y') \in R$.
            \end{itemize}
        \item ($F = \nondet_{\Sigma}$).
            The $(\nondet_{\Sigma}, \rellift{\nondet_{\Sigma}})$-bisimilarity $\nu(\pull{c} \circ \rellift{\nondet_{\Sigma}})$
            for an $\nondet_{\Sigma}$-coalgebra $c \colon C \to \nondet_{\Sigma} C$
            is the ordinary bisimilarity for the NA $c$,
            that is the maximum relation $R$ on $C$ such that
            if $(x, y) \in R$ then $\pi_1 (c(x)) = \pi_1 (c(y))$ and
            \begin{itemize}
                \item for each $a \in \Sigma, x' \in \pi_2(c(x))(a)$,
                    there is $y' \in \pi_2(c(y))(a)$ such that $(x' , y') \in R$, and
                \item for each $a \in \Sigma$ and $y' \in \pi_2(c(y))(a)$,
                    there is $x' \in \pi_2(c(x))(a)$ such that $(x', y') \in R$.
            \end{itemize}
        \item ($F = \distr$) \cite{deVinkRutten99}.
            The $(\distr, \rellift{\distr})$-bisimilarity $\nu(\pull{c} \circ \rellift{\distr})$
            for a $\distr$-coalgebra $c \colon C \to \distr C$ is
            the maximum relation $R$ such that
            if $(x,y) \in R$ then
            $\sum_{z \in K} c(x)(z) = \sum_{z \in K} c(y)(z)$
            for every equivalence class $K \subseteq C$ of $R$.
    \end{enumerate}
\end{myexample}

\section{Fibrational Partitioning}\label{sec:fibrational-partitioning}
We introduce the notion of fibrational partitioning, one that is central to our algorithm that grows a tree using fibre objects (cf.\ \cref{fig:Partitioning-intro}).

Given an ``equivalence relation'' $R$ over $C$---identified with an object $R\in \cate_{C}$ over $C\in \catc$ in a suitable fibration $p \colon \cate \to \catc$---a fibrational $R$-partitioning is a mono-sink, shown on the right, 
that is subject to certain axioms.
The notion allows us to explicate equivalence classes (namely $\{C_{i}\}_{i}$) in the abstract fibrational language. 
\begin{equation*}
    \begin{tikzpicture}[x=1.5em,y=2em]
        \node (E) at (4, 1.5) {$\cate$} ;
        \node (C) at (4, 0) {$\catc$} ;
        \draw[->] (E) -- node[right]{$p$} (C) ;
        \node (R) at (3, 1.5) {$R$} ;
        \node (C0)   at (0, 0.8) {$C_0$} ;
        \node (dots) at (0, 0) {$\vdots$} ;
        \node (Cn)   at (0, -0.8) {$C_n$} ;
        \node (C)    at (3, 0) {$C$} ;
        \draw[>->] (C0) -- node[above]{$\kappa_{0}$} (C) ;
        \draw[>->] (Cn) -- node[below]{$\kappa_{n}$} (C) ;
        \draw[|->] (R) -- (C) ;
    \end{tikzpicture}
\end{equation*}

\begin{mydefinition}[$R$-partitioning]
	\label{def:R-partitioning}
	Let $\catc$ be a category with pullbacks and an initial object $0$,
  and $p \colon \cate \to \catc$ be a $\clat$-fibration.
  Let $C\in \catc$ and $R\in \cate_{C}$. An \emph{$R$-partitioning} is a mono-sink (i.e.\ a family of monomorphisms)
  $\{ \kappa_i \colon C_i \monoto C \}_{i \in I}$ that satisfies:
  \begin{enumerate}
      \item\label{item:R-part-cond-1} $\pull{\kappa_i}(R) = \top_{C_i}$ for all $i \in I$,
      \item\label{item:R-part-cond-2} $\bigsqcup_{i \in I} \push{(\kappa_i)}(\top_{C_i}) = R$, and
      \item\label{item:R-part-cond-3} $C_i \not\cong 0$ and $C_i \cap C_j \cong 0$ for each $i, j \in I$ with $i \ne j$.
  \end{enumerate}
  We say a $\clat$-fibration  $p$ \emph{admits partitioning} if (1) for each $C\in \catc$ and $R \in \cate_{C}$, there is an $R$-partitioning; 
  and moreover, (2) for each $C \in \catc$, $R, R' \in \cate_{C}$ such that $R' \sqsubseteq R$,
  and each $R$-partitioning $\{ \kappa_i \colon C_i \monoto C \}_{i \in I}$,
  we have $\textstyle\bigsqcup_{i\in I} \push{(\kappa_i)} (\pull{\kappa_i} R') = R'$.
\end{mydefinition}
Cond.~\itemref{item:R-part-cond-3} asserts that the components $C_{i}$ are nontrivial and disjoint.
Cond.~\itemref{item:R-part-cond-1} says the partitioning $\{C_{i}\}_{i}$ is \emph{not too coarse}---the original equivalence $R$,
when restricted to $C_{i}$, should relate all pairs of elements in $C_i$.
Conversely, Cond.~\itemref{item:R-part-cond-2} means that  $\{C_{i}\}_{i}$ is \emph{not too fine}---if it were finer than $R$,
then the relation $\bigsqcup_{i \in I} \push{(\kappa_i)}(\top_{C_i})$ over $C$ would be finer than $R$.
See the concrete description of $\push{(\kappa_i)}$ in~\cref{ex:eqrelIsACLatFib}. 

\begin{myexample}[$\eqrel \to \sets$ admits partitioning]
    $\eqrel \to \sets$ admits partitioning. Indeed, given an equivalence relation $R\in \eqrel_{C}$ over $C$, the mono-sink
        $\{\kappa_{S}\colon S\monoto C\}_{S\in C/R}$, 
        where  $S\in C/R$ is naturally identified with a subset of $C$,
    is an $R$-partitioning.
    Cond.~\itemref{item:R-part-cond-1}--\itemref{item:R-part-cond-3} are easily verified following Example~\ref{ex:eqrelIsACLatFib}. 

    An $R$-partitioning is not necessarily unique.
    This happens when $R\in \eqrel_{C}$ has singleton equivalence classes.
    Let $A\subseteq C$ be an arbitrary subset such that each $x\in A$ composes a singleton $R$-equivalence class.
    Then
      $\{\kappa'_{S}\colon S\monoto C\}_{S\in I}$, 
      where $I=(C/R)\setminus\bigl\{\, \{x\} \,\big|\, x \in A \,\bigr\}$,
    is also an $R$-partitioning.
 With this mono-sink (that is ``narrower'' than the original
 $\{\kappa_{S}\}_{S\in C/R}$),
    Cond.~\ref{item:R-part-cond-2} is satisfied since the equivalence closure operation included in the direct images $\push{(\kappa_i)}(\top_{C_i})$
    (see Example~\ref{ex:eqrelIsACLatFib}) compensates the absence of $x\in A$.
\end{myexample}

    \begin{figure}[t]
        \centering
        \footnotesize
        \begin{minipage}{0.49\linewidth}
            \centering
            (Beck--Chevalley)
    
            \noindent
            \begin{math}
                \begin{tikzcd}[ampersand replacement=\&, row sep=small, column sep=small]
                    A \cap B
                    \ar[d, >->, "m"']
                    \ar[r, >->, "n"]
                    \ar[rd, phantom, "\usebox\pullback", very near start]
                    \&
                    B
                    \ar[d, >->, "\lambda"]
                    \\
                    A
                    \ar[r, >->, "\kappa"]
                    \&
                    C
                \end{tikzcd}
                \;\Longrightarrow\;
                \begin{tikzcd}[ampersand replacement=\&, row sep=small]
                        \cate_{A \cap B}
                        \ar[d, "\push{m}"']
                        \&
                        \cate_{B}
                        \ar[l, "\pull{n}"']
                        \ar[d, "\push{\lambda}"]
                        \\
                        \cate_{A}
                        \&
                        \cate_{C}
                        \ar[l, "\pull{\kappa}"]
                \end{tikzcd}
            \end{math}
        \end{minipage}
        \begin{minipage}{0.49\linewidth}
            \centering
            (modularity)
    
            \noindent
            \begin{math}
                \begin{tikzcd}[ampersand replacement=\&, row sep=small, column sep=small]
                    \cate_A \times \cate_B
                    \ar[r, "{\push{\kappa} \times \push{\lambda}}"]
                    \&
                    \cate_C \times \cate_C
                    \ar[r, "\sqcup"]
                    \ar[d, "\pull{\lambda} \times \pull{\lambda}"']
                    \&
                    \cate_C
                    \ar[d, "\pull{\lambda}"]
                    \\
                    \&
                    \cate_{B} \times \cate_{B}
                    \ar[r, "\sqcup"]
                    \&
                    \cate_{B}
                \end{tikzcd}
            \end{math}
        \end{minipage}
        \caption{Conditions for \cref{assum:well-compatible-fib}.}
        \label{fig:well-compatible-fib}
    \end{figure}
The fibration $\eqrel \to \sets$ is our leading example,
and unfortunately, the only example that we know admits partitioning.
There are many other examples of $\clat$-fibrations (see~\cite{Komorida+22}), but
they fail to admit partitioning, typically due to the failure of 
Cond.~\ref{item:R-part-cond-2} of \cref{def:R-partitioning}.
This absence of examples does not harm the value of our fibrational framework:
our goal is to explicate categorical essences of partition refinement;
and we do not aim at new instances via categorical abstraction (although such are certainly desirable).

\section{The Naive Fibrational Algorithm $\PRNaive$}\label{sec:naiveAlgo}
We introduce a naive fibrational partition refinement algorithm, called $\PRNaive$,
as a preparation step to our main algorithm $\PRHopcroft$ (\cref{algo:fibrational-block-specified}).

In what follows, a prefix-closed set $T \subseteq \nat^{*}$ (where $\nat^*$ is the set of strings over $\nat$) is identified with a rooted tree.
We denote the leaves of $T$ by $\leaves(T)$.

We introduce further conditions that make fibrations ``compatible'' with partitioning.
It is easy to see that $\eqrel \to \sets$ satisfies the conditions on $p$ in \cref{assum:well-compatible-fib}.
\begin{myassumption}\label{assum:well-compatible-fib}
    Assume a $\clat$-fibration  $p \colon \cate \to \catc$ that satisfies the following conditions. 
    \begin{enumerate}
        \item\label{item:good-fib-distributive} For each $C \in \catc$,
            the lattice $\sub(\catc)_C$ of subobjects of $C$ in $\catc$ is distributive.
        \item\label{item:good-fib-no-interference} (Beck--Chevalley) For every
	        pullback diagram along monomorphisms in $\catc$, shown in 
            the first diagram in \cref{fig:well-compatible-fib},
            the induced diagram, the second in \cref{fig:well-compatible-fib}, commutes. 
        \item\label{item:good-fib-fibre-modular} For any monomorphisms $\kappa \colon A \monoto C$ and $\lambda \colon B \monoto C$,
            the third diagram in \cref{fig:well-compatible-fib} is a fork, where the following diagram is a \emph{fork} if
            $h_1 \circ g_1 \circ f = h_2 \circ g_2 \circ f$.
            \begin{equation*}
                \begin{tikzcd}[row sep=small]
                    W \ar[r, "f"] & X \ar[r, "g_1"] \ar[d, "g_2"'] & Y_1 \ar[d, "h_1"]\\
                    & Y_2 \ar[r, "h_2"] & Z
                \end{tikzcd}
            \end{equation*}
    \end{enumerate}
\end{myassumption}

\begin{mydefinition} \label{def:naive-algorithm}
    Let $p \colon \cate \to \catc$ be a $\clat$-fibration that satisfies \cref{assum:well-compatible-fib},
    $F \colon \catc \to \catc$ be a functor,
    and  $\lift{F} \colon \cate \to \cate$ be its lifting along $p$ (\cref{def:lifting}).
    \cref{algo:naive} shows our \emph{naive fibrational partition refinement algorithm}.
    Given a coalgebra $c \colon C \to FC$, 
    it computes a $\nu(\pull{c}\lift{F})$-partitioning of $C$,
    i.e.\ modulo the $(F,\lift{F})$-bisimilarity of $c$ (\cref{def:fibrational-bisim}).
\end{mydefinition}

\begin{algorithm}[htbp]
    \caption{The naive fibrational partition refinement algorithm $\PRNaive$.}
    \label{algo:naive}
    \begin{algorithmic}[1]
        \Require A coalgebra $c \colon C \to FC$ in $\catc$.
        \Ensure A mono-sink $\{ \kappa_i \colon C_i \monoto C \}_{i \in I}$ for some $I$.
        \State $T := \{ \epsilon \} \subseteq \nat^{*}$;
        $C_{\epsilon} := C$;
        $\kappa_{\epsilon} := \idmorph_{C} \colon C_{\epsilon} \monoto C$;
        $R := \top_C$
        \Comment{initialisation}
        \While{$\pull{c}\lift{F}R \ne R$} \label{line:naive-main-loop}\Comment{the main loop}
            \State\label{line:naive-update-R}$R := \pull{c}\lift{F}R$; $L := \leaves(T)$
            \For{$\sigma \in L$}
                \State Take a $\pull{\kappa_{\sigma}}(R)$-partitioning $\{ \lambda_{\sigma, k} \colon C_{\sigma k} \monoto C_{\sigma}\}_{k \in \{ 0, \dots, n_{\sigma} \}}$ of $C_{\sigma}$
                \For{$k = 0, \dots, n_{\sigma}$}
                    $\kappa_{\sigma k} := \kappa_{\sigma} \circ \lambda_{\sigma, k} \colon C_{\sigma k} \monoto C$
                \EndFor
                \State $T := T \cup \{ \sigma 0, \dots, \sigma n_{\sigma}\}$
            \EndFor
        \EndWhile
        \State \Return $\{ \kappa_{\sigma} \colon C_{\sigma} \monoto C\}_{\sigma \in \leaves(T)}$
    \end{algorithmic}
\end{algorithm}

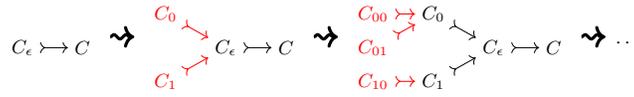
\begin{figure}[h]
    \centering
    \scalebox{0.8}{
    \begin{tikzpicture}[baseline=0]
        \node (C)  at (0,  0) {$C$} ;
        \node (Ce) at (-1, 0) {$C_{\epsilon}$} ;
        \draw[>->] (Ce) to (C) ;
    \end{tikzpicture}
    $\;\begin{tikzpicture}[x=4mm, segment length=2mm]
        \draw[->, decorate, decoration={snake,amplitude=1pt}, line width=1.5pt] (0,0) to (1,0) ;
    \end{tikzpicture}\;$
    \begin{tikzpicture}[baseline=0, yscale=0.8]
        \node (C)  at (0,  0) {$C$} ;
        \node (Ce) at (-1, 0) {$C_{\epsilon}$} ;
        \node[red] (C0) at (-2, 0.7)  {$C_0$} ;
        \node[red] (C1) at (-2, -0.7) {$C_1$} ;
        \draw[>->] (Ce) to (C) ;
        \draw[>->, red] (C0) to (Ce) ;
        \draw[>->, red] (C1) to (Ce) ;
    \end{tikzpicture}
    $\;\begin{tikzpicture}[x=4mm, segment length=2mm]
        \draw[->, decorate, decoration={snake,amplitude=1pt}, line width=1.5pt] (0,0) to (1,0) ;
    \end{tikzpicture}\;$
    \begin{tikzpicture}[baseline=0, yscale=0.8]
        \node (C)  at (0,  0) {$C$} ;
        \node (Ce) at (-1, 0) {$C_{\epsilon}$} ;
        \node (C0) at (-2, 0.7)  {$C_0$} ;
        \node (C1) at (-2, -0.7) {$C_1$} ;
        \node[red] (C00) at (-3, 0.7) {$C_{00}$} ;
        \node[red] (C01) at (-3, 0) {$C_{01}$} ;
        \node[red] (C10) at (-3, -0.7) {$C_{10}$} ;
        \draw[>->] (Ce) to (C) ;
        \draw[>->] (C0) to (Ce) ;
        \draw[>->] (C1) to (Ce) ;
        \draw[>->, red] (C00) to (C0) ;
        \draw[>->, red] (C01) to (C0) ;
        \draw[>->, red] (C10) to (C1) ;
    \end{tikzpicture}
    $\;\begin{tikzpicture}[x=4mm, segment length=2mm]
        \draw[->, decorate, decoration={snake,amplitude=1pt}, line width=1.5pt] (0,0) to (1,0) ;
    \end{tikzpicture}\; \cdots$
    }
    \caption{The $R$-partitioning gets finer as the algorithm runs.}
    \label{fig:naive-tree}
\end{figure}
\cref{algo:naive} starts with $R = \top_C \in \cate_C$
and a singleton family of a monomorphism $\{ \kappa_{\epsilon} \colon C_{\epsilon} \monoto C \}$.
With each iteration, the object $R$ on $C$ gets smaller and closer to $\nu(\pull{c}\lift{F})$
and $R$-partitioning $\{ \kappa_{\sigma} \colon C_{\sigma} \monoto C \}_{\sigma}$ gets finer (see \cref{fig:naive-tree}).
When the algorithm terminates, $R$ is equal to $\nu(\pull{c}\lift{F})$ and
a $\nu(\pull{c}\lift{F})$-partitioning is returned.

Combining the loop invariant (\cref{lem:naive-invariant}) and termination (\cref{lem:naive-termination}),
we can prove the correctness of the naive algorithm.

 \begin{mylemmarep}[loop invariant]\label{lem:naive-invariant}
    At the beginning of each iteration of the main loop,
    the following hold.
    \begin{enumerate}
        \item\label{item:naive-inv-part} The mono-sink $\{ \kappa_{\sigma} \colon C_{\sigma} \monoto C \}_{\sigma \in \leaves(T)}$ is an $R$-partitioning.
        \item\label{item:naive-inv-R} $\nu(\pull{c}\lift{F}) \sqsubseteq R$.
    \end{enumerate}
 \end{mylemmarep}
 \begin{proof}
    Firstly we prove the following two lemmas.

    \begin{mylemmarep}\label{lem:well-compatible-fib}
        Assume a $\clat$-fibration  $p \colon \cate \to \catc$ satisfies \cref{assum:well-compatible-fib}.
        Now let 
        $\{ \kappa_i \colon A_i \monoto C \}_{i \in I}$ be a mono-sink in $\catc$ that is pairwise disjoint,
        $\lambda \colon B \monoto C$ be a monomorphism with $A_i \cap B \cong 0$ for each $i \in I$, $P_i \in \cate_{A_i}$,
        and $R \in \cate_B$. Then we have
        $\pull{\lambda} \left( \left( \textstyle\bigsqcup_{i \in I} \push{(\kappa_i)}(P_i) \right) \sqcup \push{\lambda}(R) \right) = R$.
    \end{mylemmarep}   

    The above claim can be understood as follows: 
    when a monomorphism $B \monoto C$ is added to a mono-sink $\{\kappa_i \colon A_i \monoto C\}_{i \in I}$,
    if $B$ is disjoint from each $A_{i}$ ($A_i \cap B \cong 0$),
    then the objects $P_{i}$ above $A_{i}$ do not interfere with $R$ above $B$.
    \begin{equation*}
        \begin{tikzpicture}[baseline=1cm, scale=0.7]
            \node (cate) at (3.5, 2) {$\cate$} ;
            \node (catc) at (3.5, 0) {$\catc$} ;
            \draw[->] (cate) -- node[right]{$p$} (catc) ;
            \node (Pudots) at (-1.4, 2.4) {$\ddots$} ;
            \node (Pi)     at (-1, 1.9) {$P_i$} ;
            \node (Pddots) at (-0.6, 1.6) {$\ddots$} ;
            \node (R)      at (0.5, 1.3) {$R$} ;
            \node (lamR)   at (2.5, 1.3) {$\push{\lambda}R$} ;
            \node (kapPi)  at (2.5, 1.9) {$\push{(\kappa_i)}P_i$} ;
            \draw[|->] (R)  -- node[below] {$\push{\lambda}$} (lamR) ;
            \draw[|->] (Pi) -- node[above] {$\push{(\kappa_i)}$} (kapPi) ;
            \node (Audots) at (-1.4, 0.4) {$\ddots$} ;
            \node (Ai)     at (-1, -0.1) {$A_i$} ;
            \node (Addots) at (-0.6, -0.4) {$\ddots$} ;
            \node (B)      at (0.5, -0.7) {$B$} ;
            \node (C)      at (2.5, 0) {$C$} ;
            \draw[>->] (B) -- node[below]{$\lambda$} (C) ;
            \draw[>->] (Ai) -- node[above]{$\kappa_i$} (C) ;
        \end{tikzpicture}
    \end{equation*}
    
    Cond.~\itemref{item:good-fib-fibre-modular} can be equationally expressed by
    \begin{math}
        \pull{\lambda} (\push{\kappa}(Q) \sqcup \push{\lambda}(R)) = (\pull{\lambda} \push{\kappa}(Q)) \sqcup (\pull{\lambda} \push{\lambda}(R))
    \end{math};
    it can be thought of as a fibrational analogue of \emph{modularity} of a lattice. 
    
    \begin{inlineproof}
        Let $A = \bigcup_{i \in I} A_i$, $\kappa \colon A \monoto C$,
        and $\iota_i \colon A_i \monoto A$ for each $i \in I$.
        Notice that the following diagram is a pullback
        since $\sub(\catc)_C$ is distributive (Cond.~\itemref{item:good-fib-distributive}):
        \begin{equation}\label{diag:empty-intersection}
            \begin{tikzcd}
                0 = B \cap (\bigcup_{i \in I} A_i)
                \ar[d, >->]
                \ar[r, >->]
                \ar[rd, phantom, "\usebox\pullback", very near start]
                &
                B
                \ar[d, >->, "\lambda"]
                \\
                \bigcup_{i \in I} A_i
                \ar[r, >->, "\kappa"]
                &
                C\mathrlap{\text{.}}
            \end{tikzcd}
        \end{equation}
        We have
        \begin{align*}
            & \pull{\lambda} \left(
                \left( \bigsqcup_{i \in I} \push{(\kappa_i)}(P_i) \right)
                \sqcup \push{\lambda}(R)
            \right)
            \\
            & =
            \pull{\lambda} \left(
                \left( \bigsqcup_{i \in I} \push{(\kappa \circ \iota_i)}(P_i) \right)
                \sqcup \push{\lambda}(R)
            \right)
            \\
            & = 
            \pull{\lambda} \left(
                \push{\kappa} \left( \bigsqcup_{i \in I} \push{(\iota_i)}(P_i) \right)
                \sqcup \push{\lambda}(R)
            \right)
            & \text{$\push{\kappa}$: left adjoint}
            \\
            & =
            \left( \pull{\lambda}  \push{\kappa} \left( \bigsqcup_{i \in I} \push{(\iota_i)}(P_i) \right) \right)
            \sqcup
            \left( \pull{\lambda} \push{\lambda}(R) \right)
            & \text{Cond.~\itemref{item:good-fib-fibre-modular}}
            \\
            & = \bot \sqcup R
            & \text{\cref{diag:empty-intersection} and Cond.~\itemref{item:good-fib-no-interference}}
            \\
            & = R.
        \end{align*}
    \end{inlineproof}

    The following lemma intuitively says that a partitioning can be refined by another partitioning of one of its leaves, in the manner shown as follows.
    \begin{equation*}
        \begin{tikzpicture}
            \node (C) at (0, 0) {$C$};
            \node (C0)    at (-1.4, 1) {$C_0$};
            \node (C1)    at (-1.4, 0.5) {$C_1$};
            \node (Cdots) at (-1.4, 0) {$\vdots$};
            \node (Cm)    at (-1.4, -0.5) {$C_m$};
            \draw[>->] (C0) -- node[above]{$\kappa_0$} (C) ;
            \draw[>->] (C1) -- node[below]{$\kappa_1$} (C) ;
            \draw[>->] (Cm) -- node[below]{$\kappa_m$} (C) ;
            \node (D0)    at (-2.8, 1.5) {$D_0$} ;
            \node (Ddots) at (-2.8, 1) {$\vdots$} ;
            \node (Dn)    at (-2.8, 0.5) {$D_n$} ;
            \draw[>->] (D0) -- node[above]{$\lambda_0$} (C0) ;
            \draw[>->] (Dn) -- node[below]{$\lambda_n$} (C0) ;
        \end{tikzpicture}
    \end{equation*}
    \begin{mylemmarep}\label{lem:part-part}
        Let $p \colon \cate \to \catc$ be a $\clat$-fibration that satisfies the three conditions in \cref{lem:well-compatible-fib}, $C\in \catc$,
        $R\in \cate_{C}$,
        $\{ \kappa_{j} \colon C_{j} \monoto C\}_{j \in \{0, \dots, m\}} $ be an $R$-partitioning of $C$,
        $R_0$ be an object of $\cate_{C_0}$,
        and $\{ \lambda_{k} \colon D_{k} \monoto C_0 \}_{k \in \{0, \dots, n\}}$ be an $R_0$-partitioning of $C_0$.
        The family of monomorphisms
        \begin{equation*}
        \textstyle
            \Pi = \left\{ \kappa_0 \circ \lambda_k \colon D_k \monoto C \right\}_{k \in \{0,\dots , n\} }
            \cup \left\{ \kappa_j \colon C_j \monoto C \right\}_{j \in \{1, \dots, m\}}
        \end{equation*}
        is a $Q$-partitioning of $C$ where
        \begin{equation*}
            \begin{aligned}
            Q & =
            \bigl( \bigsqcup_{k = 0}^n {\push{(\kappa_0 \circ \lambda_k)}(\top_{D_k})} \bigr)
            \sqcup
            \left( \bigsqcup_{j = 1}^m {\push{(\kappa_j )}(\top_{C_j})} \right) \\
            & =
            \push{(\kappa_0)}(R_0)
            \sqcup
            \left( \bigsqcup_{j = 1}^m {\push{(\kappa_j )}(\top_{C_j})} \right).
            \end{aligned}
        \end{equation*}
    \end{mylemmarep}
    \begin{inlineproof}
        We check that $\Pi$ satisfies the conditions \itemref{item:R-part-cond-1}--\itemref{item:R-part-cond-3} of \cref{def:R-partitioning}.

        $\Pi$ satisfies \cref{def:R-partitioning}.\itemref{item:R-part-cond-2} by the definition of $Q$.

        For each $k \in \{ 0, \dots , n\}$ and $j \in \{ 1 , \dots, m \}$,
        we have $D_k \cap C_j = 0$.
        Since $\{ \kappa_{j} \colon C_{j} \monoto C\}_{j \in \{0, \dots, m\}} $ and
        $\{ \lambda_{k} \colon D_{k} \monoto C_0 \}_{k \in \{0, \dots, n\}}$
        are partitioning, we have $C_j \cap C_{j'} = 0$ for $j, j' \in \{ 1, \dots, m \}$ with $j \ne j'$
        and $D_k \cap D_{k'} = 0$ for $k, k' \in \{ 1, \dots, n \}$ with $k \ne k'$.
        Moreover, we have $C_j \not\cong 0$ and $D_k \not\cong 0$ for each $j$ and $k$.
        Thus, $\Pi$ satisfies \cref{def:R-partitioning}.\itemref{item:R-part-cond-3}.

        We have
        \begin{align*}
            \pull{\kappa_j}(Q)
            & = \pull{\kappa_j}\left(
                \push{(\kappa_0)}(R_0)
                \sqcup
                \left( \bigsqcup_{j' = 1}^m {\push{(\kappa_{j'} )}(\top_{C_{j'}})} \right)
                \right)
            \\
            & = \top_{C_j}
            & \text{\cref{lem:well-compatible-fib}}
        \end{align*}
        for $j \in \{ 1, \dots, m \}$,
        and
        \begin{align*}
            & \pull{(\kappa_0 \circ \lambda_k)}(Q)
            \\
            & = \pull{(\kappa_0 \circ \lambda_k)}\left(
                \left( \bigsqcup_{k' = 0}^n {\push{(\kappa_0 \circ \lambda_{k'})}(\top_{D_{k'}})} \right)
                \sqcup
                \left( \bigsqcup_{j = 1}^m {\push{(\kappa_j )}(\top_{C_j})} \right)
                \right)
            \\
            & = \top_{D_k}
            & \text{\cref{lem:well-compatible-fib}}
        \end{align*}
        for $k \in \{ 0, \dots, n \}$.
        Hence, $\Pi$ satisfies \cref{def:R-partitioning}.\itemref{item:R-part-cond-1}.
        Therefore, $\Pi$ is a $Q$-partitioning of $C$.
    \end{inlineproof}

    We go back to the proof of \cref{lem:naive-invariant}.
    We write $R_i$ and $T_i$ for $R$ and $T$, respectively, at the beginning (line \ref{line:naive-main-loop}) of the $i$-th iteration of the main loop.
    We prove \itemref{item:naive-inv-part} and \itemref{item:naive-inv-R} by the induction on $i$.
    
    (\textbf{Base case}).
    We have $R_0 = \top_C$ and $T_0 = \{ \epsilon \}$.
    The mono-sink is $\{ \kappa_{\epsilon} \colon C_{\epsilon} \monoto C \}$, and this is an $R_0$-partitioning of $C$.
    We also have $\nu(\pull{c}\lift{F}) \sqsubseteq \top_C = R_0$.

    (\textbf{Induction step}).
    Assume that $\{ \kappa_{\sigma} \colon C_{\sigma} \monoto \}_{\sigma \in \leaves(T_i)}$ is an $R_i$-partitioning
    and $\nu(\pull{c}\lift{F}) \sqsubseteq R_i$.
    By $\nu(\pull{c}\lift{F}) \sqsubseteq R_i$,
    we have $\nu(\pull{c}\lift{F}) = \pull{c}\lift{F}(\nu(\pull{c}\lift{F})) \sqsubseteq \pull{c}\lift{F}(R_i) = R_{i + 1}$.
    Hence, \itemref{item:naive-inv-R} holds for $i + 1$.
    By \cref{lem:part-part} and the induction hypothesis, we have
    $\{ \kappa_{\sigma} \colon C_{\sigma} \monoto C \}_{\sigma \in \leaves(T_{i + 1})}$ is an $R_{i + 1}$-partitioning.
    Therefore, \itemref{item:naive-inv-part} holds for $i + 1$.
 \end{proof}

\begin{mylemmarep}[termination]\label{lem:naive-termination}
  If $\cate_C$ is a well-founded lattice, \cref{algo:naive} terminates.
\end{mylemmarep}
\begin{proof}
  We write $R_i$ for $R$ at the beginning (line \ref{line:naive-main-loop}) of the $i$-th iteration of the main loop,
  and have
  \begin{equation*}
      R_0 = \top_{C}, \quad R_1 = \pull{c}\lift{F} R_0, \quad \dots, \quad R_{i + 1} = \pull{c}\lift{F} R_i, \quad \dots.
  \end{equation*}
  We have $R_1 \sqsubseteq \top_C = R_0$.
  It is proven by the induction on $i$ and the functoriality of $\pull{c}\lift{F}$ that
  $R_{i} = \pull{c}\lift{F} R_{i - 1} \sqsubseteq R_{i - 1}$ holds for each $i$ such that $R_i$ is defined.
  By this observation and the condition in line \ref{line:naive-main-loop}, we have the strictly descending sequence in $\cate_C$:
  $ R_0 \sqsupset R_1 \sqsupset \cdots \sqsupset R_n \sqsupset \cdots$.
  Since $\cate_C$ is well-founded, the length of the sequence is finite,
  that is \cref{algo:naive} terminates.
\end{proof}

\begin{mypropositionrep}[correctness] \label{prop:naive-algorithm-correct}
  If $\cate_C$ is well-founded, then
  \cref{algo:naive} terminates
  and returns $\nu(\pull{c}\lift{F})$-partitioning $\{\kappa \colon C_{i} \monoto C \}_{i \in I}$.
\end{mypropositionrep}
\begin{proof}
  By \cref{lem:naive-termination}, \cref{algo:naive} terminates.
  From the termination condition (line \ref{line:naive-main-loop}) of the main loop,
  $R = \pull{c}\lift{F}R$ holds when the algorithm terminates.
  Hence, $R$ is a fixed point.
  By \cref{lem:naive-invariant}.\itemref{item:naive-inv-R},
  $R$ is greater than or equal to the greatest fixed point $\nu(\pull{c}\lift{F})$ of the functor $\pull{c}\lift{F}$.
  Thus, we have $R = \nu(\pull{c}\lift{F})$.
\end{proof}

\section{Optimised Algorithms with Hopcroft's Inequality}\label{sec:optimisedAlgorithms}
Recall that the naive algorithm grows a tree \emph{uniformly} so that every
leaf has the same depth (see \cref{fig:naive-tree}; note that, even if $C_{\sigma}$ is fine enough, we extend the node by a trivial partitioning).
By selecting leaves in a smart way and generating a tree selectively, the time cost of each iteration can be
reduced, so that Hopcroft's inequality is applicable.

In \cref{subsec:PRHopcroft}, we present a  functor-generic and fibrational algorithm enhanced with the Hopcroft-type optimisation,
calling it $\PRHopcroft$. 
We  use Hopcroft's inequality (\cref{sec:Hopcroft-trick}) for  complexity analysis.

In \cref{sec:complexity-analysis} we instantiate $\PRHopcroft$ to the fibration $\eqrel \to \sets$, obtaining three concrete (yet functor-generic) algorithms 
$
\PRHopcroftEqRel{\wCard},
\PRHopcroftEqRel{\wPred},
\PRHopcroftEqRel{\wReach}
$ that use different weight functions.
$\PRHopcroftEqRel{\wCard}$ is essentially the algorithm in~\cite{JacobsWissmann23}.
The other two ($\PRHopcroftEqRel{\wPred}, \PRHopcroftEqRel{\wReach}$) use the weight functions from the works~\cite{Hopcroft71, Gries73, Knuutila01} on DFA partition refinement.
The three algorithms exhibit slightly different asymptotic complexities.

\subsection{A Fibrational Algorithm $\PRHopcroft$ Enhanced by  Hopcroft's Inequality }
\label{subsec:PRHopcroft}
We fix a $\clat$-fibration $p \colon \cate \to \catc$, 
functors $F \colon \catc \to \catc$ and $\lift{F} \colon \cate \to \cate$,
an $F$-coalgebra $c \colon C \to FC$, and
a map $w \colon \obj(\sub(\catc)_C) \to \nat$ (which we use for weights).
We write $w(C')$ for $w(\lambda \colon C' \monoto C)$ when no confusion arises.

The following conditions clarify which properties of $\eqrel \to \sets$ are necessary to make
our optimised fibrational algorithm $\PRHopcroft$ work:
the last one (\cref{assum}.\itemref{assum:weight}) is for complexity analysis;
all the other ones are for correctness. 
\begin{myassumption}\label{assum}
    \begin{enumerate}
        \item \label{assum:C-nice} $\catc$ has pullbacks, pushouts along monos, and an initial object $0$.
        \item \label{assum:fibre-0} The fibre category $\cate_0$ above an initial object $0$ is trivial, that is $\top_0 = \bot_0$.
        \item \label{assum:fibred} $\lift{F}$ is a fibred lifting of $F$ along $p$.
        \item \label{assum:F-preserves-mono} $F\colon \catc \to \catc$ preserves monomorphisms whose codomain is not $0$.
        \item \label{assum:monotone-partitioning} The fibration $p$ admits partitioning.
        \item \label{assum:good-fibration} The fibration $p \colon \cate \to \catc$ satisfies the three conditions in \cref{assum:well-compatible-fib}.
        \item \label{assum:well-founded} The fibre category $\cate_{C}$ is a well-founded lattice.
        \item \label{assum:finite-partitioning} If $C' \monoto C$ and $R \in \cate_{C'}$,
            every $R$-partitioning $\{ \lambda_k \colon D_k \monoto C' \}_{k \in K}$ is finite ($|K| < \infty$).
        \item \label{assum:inj-on-obj} If $\kappa \colon A \monoto C$ and $\lambda \colon B \monoto C$ are monomorphisms and $A \cap B \cong 0$,
            then the functor
            \begin{tikzcd} \cate_A \times \cate_B \ar[r, "{\push{\kappa} \times \push{\lambda}}"] & \cate_C \times \cate_C \ar[r, "\sqcup"] & \cate_C \end{tikzcd}
            is injective on objects.
        \item \label{assum:weight} For a monomorphism $C' \monoto C$,
            an object $R \in \cate_{C'}$,
            and an $R$-partitioning $\{ \kappa_i \colon C_{i} \monoto C' \}$ of $C'$,
            $\sum_{i = 1}^n w(C_{i}) \le w(C')$ holds.
    \end{enumerate}
\end{myassumption}

\cref{assum}.\itemref{assum:fibred} is not overly restrictive. Indeed,
the following functors on $\sets$ have a fibred lifting.
The functors described in \cref{example:F-coalg} are examples of the functor defined by (\ref{eq:functorBNF}).

\vspace{1mm}
\noindent
\begin{myproposition}
    Consider the endofunctors on $\sets$ defined by the BNF below.
    \begin{equation}\label{eq:functorBNF}
     \textstyle
        F \mathrel{::=} \idfunc \mid A \mid \coprod_{b\in B} F_b \mid \prod_{b\in B} F_b \mid \powset F\mid \distr F
    \qquad\text{    where $A$ and $B$ are sets.}
    \end{equation}
    The relation lifting $\rellift{F} \colon \eqrel \to \eqrel$ of $F$ (\cref{def:relation-lifting}) is fibred.
\end{myproposition}

\begin{mypropositionrep}
    The fibration $p \colon \eqrel \to \sets$
    with $\rellift{F}$ (\ref{eq:functorBNF})
    and a coalgebra $c \colon C \to FC$ for a finite set $C$
    satisfies the assumptions \itemref{assum:C-nice}--\itemref{assum:inj-on-obj} of \cref{assum}.
\end{mypropositionrep}
\begin{proof}
    We only prove that $p \colon \eqrel \to \sets$ satisfies
    the premise \itemref{item:good-fib-fibre-modular} of \cref{lem:well-compatible-fib}.
    The other conditions are easy to check.

    Given injections $\kappa \colon A \monoto C$ and $\lambda \colon B \monoto C$ in $\sets$
    and equivalence relations $R$ on $A$ and $S$ on $B$.
    We want to show
    $(\pull{\lambda} \push{\kappa} R) \sqcup (\pull{\lambda} \push{\kappa} S)
    = \pull{\lambda} ((\push{\kappa} R) \sqcup (\push{\lambda} S))$.
    For any $(x,y) \in \pull{\lambda} ((\push{\kappa} R) \sqcup (\push{\lambda} S))$,
    we have $(\lambda(x) , \lambda(y)) \in (\push{\kappa} R) \sqcup (\push{\lambda} S)$.
    Hence, there exist $m \in \nat$ and $z_0 , \dots, z_m \in C$ such that
    $z_0 = \lambda(x)$, $z_m = \lambda(y)$, and
    $(z_i, z_{i+1}) \in \push{\kappa}R$ or $(z_i, z_{i+1}) \in \push{\lambda}S$
    for each $i = 0, \dots, m-1$.
    We can assume that $z_i \ne z_{i+1}$
    and $z_i \in \lambda(B)$
    for each $i$ without loss of generality
    since $\lambda(x), \lambda(y) \in \lambda(B)$.
    Thus, the sequence
    $x = \lambda^{-1}(z_0), \lambda^{-1}(z_1), \dots, \lambda^{-1}(z_m) = y$ in $B$
    satisfies
    $(\lambda^{-1}(z_i), \lambda^{-1}(z_{i+1})) \in \pull{\lambda} \push{\kappa} R$ or
    $(\lambda^{-1}(z_i), \lambda^{-1}(z_{i+1})) \in \pull{\lambda} \push{\lambda} S$
    for each $i$.
    This means $(x, y) \in (\pull{\lambda} \push{\kappa} R) \sqcup (\pull{\lambda} \push{\lambda} S)$.
    Hence, $\pull{\lambda} ((\push{\kappa} R) \sqcup (\push{\lambda} S)) \sqsubseteq (\pull{\lambda} \push{\kappa} R) \sqcup (\pull{\lambda} \push{\lambda} S)$ holds.
    Conversely, we can show
    $\pull{\lambda} ((\push{\kappa} R) \sqcup (\push{\lambda} S)) \sqsupseteq (\pull{\lambda} \push{\kappa} R) \sqcup (\pull{\lambda} \push{\lambda} S)$
    by the similar argument.
\end{proof}

\begin{mydefinition}[$\PRHopcroft$] \label{def:fibrational-block-specified-algo}
  Let the $\clat$-fibration $p \colon \cate \to \catc$,
  the map $w$, 
  the functors $F \colon \catc \to \catc$ and $\lift{F} \colon \cate \to \cate$,
  and the object $C$
  satisfy \cref{assum}.
  \cref{algo:fibrational-block-specified} shows the algorithm $\PRHopcroft_{(F, \lift{F}), w}$
  ($(F, \lift{F})$ and $w$ are omitted when  clear from the context).
  Given a coalgebra $c \colon C \to FC$, 
  it computes a $\nu(\pull{c}\lift{F})$-partitioning of $C$,
  like the naive algorithm.
\end{mydefinition}

\begin{algorithm}[t]
  \caption{An optimised fibrational partition refinement algorithm $\PRHopcroft_{(F, \lift{F}), w}$.}
  \label{algo:fibrational-block-specified}
  \begin{algorithmic}[1]
      \Require A coalgebra $c \colon C \to FC$ in $\catc$.
      \Ensure A mono-sink $\{ \kappa_i \colon C_i \monoto C \}_{i \in I}$ for some $I$.
      
      \State\label{line:initialisation} $J := \{\epsilon\} \subseteq \nat^{*}$;\ $C_{\epsilon} := C$;\ $C^{\clean}_{\epsilon} := 0$
          \Comment initialisation
      \While{there is $\rho \in \leaves(J)$ such that $C^{\clean}_{\rho} \ne C_{\rho}$} \label{line:loop}
          \Comment the main loop

          \State\label{line:def-rel}$R := \bigsqcup_{\sigma \in \leaves(J)}\push{(\kappa_{\sigma})}(\top_{C_{\sigma}})$
          \Comment \textbf{Partitioning} (Line~\ref{line:def-rel}--\ref{line:partitioning-last-line})

          \State\label{line:choose-a-leaf}Choose a leaf $\rho \in \leaves(J)$ such that $C^{\clean}_{\rho} \ne C_{\rho}$
          \State\label{line:Rrho}$R_{\rho} := \pull{(c \circ \kappa_{\rho})}(\lift{F}(R))$
          \If{$R_{\rho} = \top_{C_{\rho}}$}
              \State\label{line:mark-all-states-clean}$C^{\clean}_{\rho} := C_{\rho}$
              \Continue
          \EndIf
          \State \label{line:partitioning-last-line}Take an $R_{\rho}$-partitioning $\{ \kappa_{\rho, k} \colon C_{\rho k} \monoto C_{\rho} \}_{k \in \{0, \dots, n_{\rho} \}}$

          \State\label{line:choose-an-index}Choose $k_0 \in \{0, \dots, n_\rho \}$
              s.t.\ $\displaystyle w(C_{\rho k_0}) = \max_{k \in \{ 0, \dots, n_{\rho}\}} w(C_{\rho k})$
          \Statex \Comment \textbf{Relabelling} (Line~\ref{line:choose-an-index}--\ref{line:relabelling-last-line})
          
          \State \Call{MarkDirty}{$\rho$, $k_0$}
          \State \label{line:relabelling-last-line} $J := J \cup \{\rho 0, \dots, \rho n_{\rho} \}$
      \EndWhile
      \State\Return $\{ \kappa_{\sigma} \colon C_{\sigma} \monoto C \}_{\sigma \in \leaves(J)}$ \label{line:return}

      \Statex
      \Procedure{MarkDirty}{$\rho$, $k_0$}\label{line:markdirty-first-line}
          \For{$k \in \{0, \dots, n_{\rho}\}$}
              $C^{\clean}_{\rho k} := C_{\rho k}$
          \EndFor
          \State\label{line:pullback}Let $B$ be the pullback of the diagram:
              $\begin{tikzcd}[row sep=small, column sep=small]
                  B
                  \ar[d]
                  \ar[r, >->]
                  \ar[dr, phantom, "\usebox\pullback" , very near start]
                  &
                  C
                  \ar[d, "c"]
                  \\
                  F\left( C_{\rho k_0} \cup \left( \bigcup_{\sigma \in \leaves(J) \setminus \{ \rho \} } C_\sigma \right) \right)
                  \ar[r, >->]
                  &
                  FC
              \end{tikzcd}$.
          \State \Comment{the bottom morphism is mono by \cref{assum}.\itemref{assum:F-preserves-mono}}
          \For{$\tau \in \leaves(J \cup \{\rho 0, \dots, \rho n_{\rho} \})$}
              \State\label{label:mark-clean} $C^{\clean}_{\tau} := C^{\clean}_{\tau} \cap B$
                  \Comment{states not in $B$ are marked as dirty}
          \EndFor
      \EndProcedure\label{line:markdirty-last-line}
  \end{algorithmic}
\end{algorithm}

\vspace{1mm}
The algorithm $\PRHopcroft$ exposes a tree structure to which Hopcroft's inequality applies.
\Cref{table:Hopcroft-ineq-and-blockspecified-algorithm} summarises how constructs in $\PRHopcroft$ fit \cref{sec:Hopcroft-trick}.

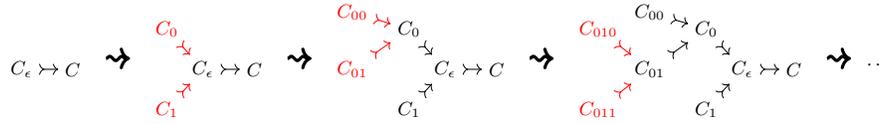
\begin{figure}[tbp]
  \centering
  \small
  \scalebox{0.8}{
  \begin{tikzpicture}[baseline=0, x=9mm, scale=0.95]
      \node (C)  at (0,  0) {$C$} ;
      \node (Ce) at (-1, 0) {$C_{\epsilon}$} ;
      \draw[>->] (Ce) to (C) ;
  \end{tikzpicture}
  $\;\;\begin{tikzpicture}[x=4mm, segment length=2mm]
      \draw[->, decorate, decoration={snake,amplitude=1pt}, line width=1.5pt] (0,0) to (1,0) ;
  \end{tikzpicture}\;\;$
  \begin{tikzpicture}[baseline=0, x=9mm, scale=0.95]
      \node (C)  at (0,  0) {$C$} ;
      \node (Ce) at (-1, 0) {$C_{\epsilon}$} ;
      \node[red] (C0) at (-1.7, 0.7)  {$C_0$} ;
      \node[red] (C1) at (-1.7, -0.7) {$C_1$} ;
      \draw[>->] (Ce) to (C) ;
      \draw[>->, red] (C0) to (Ce) ;
      \draw[>->, red] (C1) to (Ce) ;
  \end{tikzpicture}
  $\;\;\begin{tikzpicture}[x=4mm, segment length=2mm]
      \draw[->, decorate, decoration={snake,amplitude=1pt}, line width=1.5pt] (0,0) to (1,0) ;
  \end{tikzpicture}\;\;$
  \begin{tikzpicture}[baseline=0, x=9mm, scale=0.95]
      \node (C)  at (0,  0) {$C$} ;
      \node (Ce) at (-1, 0) {$C_{\epsilon}$} ;
      \node (C0) at (-1.7, 0.7)  {$C_0$} ;
      \node (C1) at (-1.7, -0.7) {$C_1$} ;
      \node[red] (C00) at (-2.8, 1) {$C_{00}$} ;
      \node[red] (C01) at (-2.8, 0) {$C_{01}$} ;
      \draw[>->] (Ce) to (C) ;
      \draw[>->] (C0) to (Ce) ;
      \draw[>->] (C1) to (Ce) ;
      \draw[>->, red] (C00) to (C0) ;
      \draw[>->, red] (C01) to (C0) ;
  \end{tikzpicture}
  $\;\;\begin{tikzpicture}[x=4mm, segment length=2mm]
      \draw[->, decorate, decoration={snake,amplitude=1pt}, line width=1.5pt] (0,0) to (1,0) ;
  \end{tikzpicture}\;\;$
  \begin{tikzpicture}[baseline=0, x=9mm, scale=0.95]
      \node (C)  at (0,  0) {$C$} ;
      \node (Ce) at (-1, 0) {$C_{\epsilon}$} ;
      \node (C0) at (-1.7, 0.7)  {$C_0$} ;
      \node (C1) at (-1.7, -0.7) {$C_1$} ;
      \node (C00) at (-2.8, 1) {$C_{00}$} ;
      \node (C01) at (-2.8, 0) {$C_{01}$} ;
      \node[red] (C010) at (-3.8, 0.7) {$C_{010}$} ;
      \node[red] (C011) at (-3.8, -0.7) {$C_{011}$} ;
      \draw[>->] (Ce) to (C) ;
      \draw[>->] (C0) to (Ce) ;
      \draw[>->] (C1) to (Ce) ;
      \draw[>->, red] (C010) to (C01) ;
      \draw[>->, red] (C011) to (C01) ;
      \draw[>->] (C00) to (C0) ;
      \draw[>->] (C01) to (C0) ;
  \end{tikzpicture}
  $\;\;\begin{tikzpicture}[x=4mm, segment length=2mm]
      \draw[->, decorate, decoration={snake,amplitude=1pt}, line width=1.5pt] (0,0) to (1,0) ;
  \end{tikzpicture}\;\;\cdots$
  }
  \caption{At each iteration one leaf of the tree is selected and refined.}
  \label{fig:block-selecting-tree}
\end{figure}

Much like $\PRNaive$ (\cref{algo:naive}),
$\PRHopcroft$ grows a tree, as shown in \cref{fig:block-selecting-tree}. We take the generated tree as $T=(V,E)$ in \cref{sec:Hopcroft-trick}. 
Note that, whereas $\PRNaive$ expands the tree uniformly so that every leaf has the same depth (\cref{fig:naive-tree}),
$\PRHopcroft$ expands leaves selectively (\cref{fig:block-selecting-tree}).

\begin{table}[t]
  \centering
  \footnotesize
  \rowcolors{2}{gray!10}{gray!0}
  \renewcommand{\arraystretch}{1.2}
  \caption{
    Correspondence between constructs in $\PRHopcroft_{(F, \lift{F}), w}$ and the theory in \cref{sec:Hopcroft-trick}
  }
  \label{table:Hopcroft-ineq-and-blockspecified-algorithm}
  \begin{tabular}{ll}
      \toprule
      Constructs in \cref{algo:fibrational-block-specified}
      &
      Notions in \cref{sec:Hopcroft-trick}
      \\
      \midrule
      The tree of subobjects $\{C_{\sigma}\}_{\sigma \in J}$ of $C$,  cf.\ \cref{fig:block-selecting-tree}
      & 
      A tree $T = (V, E)$, cf.\ \cref{fig:weighted-heavy-chosen-tree}
      \\
      A set  $\{ C_{\sigma} \}_{\sigma \in J}$ of objects in $\catc$
      &
      A set $V$ of vertices
      \\
      \begin{tabular}{l} A set $\{ \kappa_{\sigma , k} \colon C_{\sigma k} \monoto C_{\sigma}\}_{\sigma k \in J}$ \\ of monomorphisms \end{tabular}
      &
      A set  $E$  of edges
      \\
      \begin{tabular}{l} A map $w \colon \obj(\sub(\catc)_C) \to \nat$ \\ with \cref{assum}.\itemref{assum:weight} \end{tabular}
      &
      \begin{tabular}{l} A weight function \\ $w \colon V \to \nat$ (\cref{def:weight}) \end{tabular}
      \\
      A choice of $k_0$ made in Line~\ref{line:choose-an-index} of \cref{algo:fibrational-block-specified}
      &
      An hcc $h \colon V \setminus \leaves(T) \to V$ (\cref{def:heavy-child-choice})
      \\
      The complexity result of \cref{prop:time-markdirty}
      &
      The inequality of \cref{cor:time-estimation}
      \\
      \bottomrule
  \end{tabular}
  \renewcommand{\arraystretch}{1}
\end{table}

$\PRHopcroft$
chooses $k_0$ so that $w(C_{\rho k_0})$ is maximised (Line~\ref{line:choose-an-index}).
These choices constitute a  heavy child choice
(\cref{def:heavy-child-choice}), an essential construct in Hopcroft's inequality
(\cref{thm:Hopcroft-ineq}).


$\PRHopcroft$
starts with $R = \top_C \in \cate_C$,
the singleton family of a monomorphism $\{ \kappa_{\epsilon} = \idmorph_{C} \colon C_{\epsilon} \monoto C \}$, and
a marking $C^\clean_\epsilon = 0$ of states.
For each $\sigma$, $C^\clean_\sigma\monoto C_\sigma$ is a ``subset'' of $C_\sigma$ consisting of \emph{clean} states; the rest of $C_\sigma$ consists of \emph{dirty} states. Therefore, initially, all states of $C=C_\epsilon$ are marked dirty.
The main loop,
consisting of
\textbf{Partitioning}
(Line~\ref{line:def-rel}--\ref{line:partitioning-last-line})
and \textbf{Relabelling}
(Line~\ref{line:choose-an-index}--\ref{line:relabelling-last-line}),
iterates until there is no dirty state (Line~\ref{line:loop}).

The \textbf{Partitioning} part selects one leaf $C_{\rho}$ whose states include at least
one dirty state
(Line~\ref{line:choose-a-leaf}).
The tree is expanded at this selected leaf only. 
This selection makes
\cref{algo:fibrational-block-specified} different from
\cref{algo:naive}, which expands the tree at every leaf
(cf.\ \cref{fig:naive-tree} and \cref{fig:block-selecting-tree}).

The \textbf{Relabelling} part then updates the clean/dirty marking.
Firstly, it chooses one ``heavy child'' $C_{\rho k_0}$ (Line~\ref{line:choose-an-index})
from the leaves generated in \textbf{Partitioning}.
Then the iteration calls
the \textsc{MarkDirty} procedure
(Line~\ref{line:markdirty-first-line}--\ref{line:markdirty-last-line}).
It first collects states ($B$ in Line~\ref{line:pullback}) whose all ``successors'' with
respect to the coalgebra $c \colon C \to FC$ are in the object
$C_{\rho k_0} \cup \left( \bigcup_{\sigma \in \leaves(J) \setminus \{ \rho \} } C_\sigma \right)$;
the latter intuitively consists of states ``unaffected'' by tree expansion.
The procedure marks only states in $B$ as clean (Line~\ref{label:mark-clean}), which means that the rest of
the states are marked dirty.



Towards the correctness theorem of our optimised fibrational algorithm $\PRHopcroft$ (\cref{thm:correctness}),
we first make a series of preliminary observations. 

\begin{mynotation}
  We write $R_i$ for $R$ defined at Line~\ref{line:def-rel} of \cref{algo:fibrational-block-specified} at the $i$-th iteration.
  We write $J_i$ for $J$ at the beginning of the $i$-th iteration.
  We write $C^{\clean, i}_{\sigma}$ and $\kappa^{\clean, i}_{\sigma}$
  for $C^{\clean}_{\sigma}$
  and the monomorphism $\kappa^{\clean}_{\sigma} \colon C^{\clean}_{\sigma} \monoto C$, respectively,
  at  Line~\ref{line:pullback} at the $i$-th iteration.
\end{mynotation}

We identify loop invariants \cref{prop:loop-invariant}. 
Termination of $\PRHopcroft$ follows from \cref{assum}.\itemref{assum:well-founded} and \itemref{assum:inj-on-obj} (\cref{prop:termination}).
Combining these, we prove the correctness of $\PRHopcroft$ in \cref{thm:correctness}.

\begin{mypropositionrep}[loop invariants]\label{prop:loop-invariant}
  At the beginning of the $i$-th iteration,
  the following hold.
  \begin{enumerate}
      \item\label{item:clean-leaf-top} $\pull{( c \circ \kappa_{\sigma} \circ {\kappa^{\clean,i}_{\sigma}})}\lift{F}(R_i) = \top_{C^{\clean, i}_{\sigma}}$
          for each leaf $\sigma \in \leaves(J_i)$.
      \item\label{item:keep-partitioning} The mono-sink $\{ \kappa_{\sigma} \colon C_{\sigma} \monoto C \}_{\sigma \in \leaves(J_i)}$ is an $R_i$-partitioning.
      \item\label{item:gfp-R} $\nu(\pull{c}\lift{F}) \sqsubseteq R_i$.
  \end{enumerate}
  Therefore, after \cref{algo:fibrational-block-specified} terminates,
  $\pull{(c\circ \kappa_{\sigma})} \lift{F} R = \top_{C_{\sigma}}$ holds for each $\sigma \in \leaves(J)$,
  $\{ \kappa_{\sigma} \monoto C \}_{\sigma \in \leaves(J)}$ is an $R$-partitioning,
  and $\nu(\pull{c} \lift{F}) \sqsubseteq R$,
  for $R \in \cate_C$ defined in Line \ref{line:def-rel}.
\end{mypropositionrep}
\begin{proof}
  We first prove lemmas to prove the loop invariants.
  The next technical lemma follows easily from the fibredness of $\lift{F}$.
  \begin{mylemma}\label{lem:successor-and-partition}
      Let $m \colon A \to C$ be a morphism in $\catc$ and $P_1, P_2 \in \cate_C$.
      If $\pull{m} P_1 = \pull{m} P_2$,
      then, for any $C' \in \catc$,
      $\kappa \colon C' \to C$ and $\lambda \colon C' \to FA$
      such that $c \circ \kappa = Fm \circ \lambda$,
      we have
      $\pull{\kappa}(\pull{c}\lift{F}P_1) = \pull{\kappa}(\pull{c}\lift{F}P_2)$.
  \end{mylemma}
  \begin{inlineproof}
      For $j = 1, 2$, we have
      \begin{align*}
          \pull{\kappa}(\pull{c}\lift{F}P_j)
          & = \pull{\lambda}(\pull{(Fm)} \lift{F}P_j)
          & c \circ \kappa = Fm \circ \lambda
          \\
          & = \pull{\lambda}(\lift{F} (\pull{m} P_j)).
          & \text{\cref{assum}.\itemref{assum:fibred}}
      \end{align*}
      Hence,
      by the assumption $\pull{m}P_1 = \pull{m}P_2$,
      we obtain
      \[ \pull{\kappa}(\pull{c}\lift{F}P_1)
      = \pull{\lambda}(\lift{F} (\pull{m} P_1))
      = \pull{\lambda}(\lift{F} (\pull{m} P_2))
      = \pull{\kappa}(\pull{c}\lift{F} P_2).\]
  \end{inlineproof}
  
  The next lemma (\cref{lem:partition-not-changed}) identifies the subobject of $C$
  that is ``unaffected'' by the refinement from $R_{i}$ to $R_{i+1}$,
  that is, by the tree expansion at the $i$-th iteration.
  It is not hard to see that the untouched leaves in \begin{math}
  \bigcup_{\sigma \in \leaves(T) \setminus \{ \rho \} } C_\sigma 
  \end{math} belong to that ``unaffected'' part.
  A crucial observation---central to the Hopcroft-type optimisation---is that \emph{at most one} new child $C_{\rho k_0}$ can also be added,
  where we pick  the heavy one for better complexity. 
  
  \begin{mylemma}\label{lem:partition-not-changed}
      Let $\rho$ be the leaf $\rho \in \leaves(T_i) \setminus \leaves(T_{i+1})$
      that was chosen in \textbf{Partitioning} of the $i$-th iteration, and
      $k_0\in\{0,\dotsc, n_{\rho}\}$ be the  index 
    that was chosen in \textbf{Relabelling} of the $i$-th iteration.
      We have $\pull{m}(R_i) = \pull{m}(R_{i+1})$
      for the morphism
      $m \colon C_{\rho k_0} \cup \left( \bigcup_{\sigma \in \leaves(T_i) \setminus \{ \rho \} } C_\sigma \right) \monoto C$.
  \end{mylemma}
  \begin{inlineproof}
      Let $P = \bigsqcup_{\sigma \in \leaves(T_i) \setminus \{ \rho \}} \push{(\kappa_{\sigma})} (\top_{C_{\sigma}}) $,
      and $R_{\rho} = \bigsqcup_{k \in \{ 0, \dots , n_{\rho}\}} \push{(\kappa'_{\rho k})} (\top_{C_{\rho k}})$.
      By \cref{lem:well-compatible-fib}.\itemref{item:good-fib-distributive}, the following diagram is pullback:
      \begin{equation*}
          \begin{tikzcd}
              C_{\rho k_0}
              \ar[d, >->, "\iota'"']
              \ar[r, >->, "{\kappa'_{k_0}}"]
              \ar[rd, phantom, "\usebox\pullback", very near start]
              &
              C_{\rho}
              \ar[d, >->, "\kappa_{\rho}"]
              \\
              C_{\rho k_0} \cup \left( \bigcup_{\sigma \in \leaves(T_i) \setminus \{\rho\}} C_{\sigma} \right)
              \ar[r, >-> ,"m"]
              &
              C \mathrlap{\text{.}}
          \end{tikzcd}
      \end{equation*}
      Thus, the following diagram commutes by \cref{lem:well-compatible-fib}.\itemref{item:good-fib-no-interference}.
      \begin{equation}\label{diag:no-interference-for-correctness}
          \begin{tikzcd}
              \cate_{C_{\rho k_0}}
              \ar[d, "\push{\iota'}"']
              &
              \cate_{C_{\rho}}
              \ar[d, "\push{(\kappa_{\rho})}"]
              \ar[l, "\pull{(\kappa'_{k_0})}"']
              \\
              \cate_{C_{\rho k_0} \cup \left( \bigcup_{\sigma \in \leaves(T_i) \setminus \{\rho\}} C_{\sigma} \right)}
              &
              \cate_{C}
              \ar[l, "\pull{m}"]
          \end{tikzcd}
      \end{equation}
      By chasing the diagram, we have
      \begin{equation}\label{eq:mQi}
          \begin{split}
              \pull{m} (\push{(\kappa_{\rho})} (\top_{C_\rho}))
              & = \push{\iota'} (\pull{(\kappa'_{k_0})} (\top_{C_\rho}))
              \qquad \text{\cref{diag:no-interference-for-correctness}}
              \\
              & = \push{\iota'} (\top_{C_{\rho k_0}}),
              \qquad \text{$\pull{(\kappa'_{k_0})}$: right adjoint and $\top$: limit}
          \end{split}    
      \end{equation}
      and
      \begin{equation}\label{eq:mQi1}
          \begin{split}
              \pull{m} \left( \push{(\kappa_{\rho})} (R_{\rho}) \right)
              & = \push{\iota'} \left( \pull{(\kappa'_{k_0})} (R_{\rho}) \right)
              \qquad \text{\cref{diag:no-interference-for-correctness}}
              \\
              & = \push{\iota'} (\top_{C_{\rho k_0}}).
              \qquad \text{$\{ \kappa'_{\rho k} \colon C_{\rho k} \monoto C_{\rho}\}_{k}$ is $R_{\rho}$-partitioning}
          \end{split}
      \end{equation}
      There are monomorphisms
      $\iota_{\sigma} \colon C_{\sigma} \monoto C_{\rho k_0} \cup \bigcup_{\sigma \in \leaves(T_i) \setminus \{ \rho \} } C_{\sigma}$
      for each $\sigma \in \leaves(T_i) \setminus \{ \rho \}$ with $\kappa_{\sigma} = m \circ \iota_{\sigma}$.
      We have
      \begin{align*}
          & \pull{m} (R_i)
          &
          \\
          & = \pull{m} \left( \bigsqcup_{\sigma \in \leaves(T_i)} \push{(\kappa_{\sigma})} (\top_{C_{\sigma}}) \right)
          & \text{Definition of $R_i$}
          \\
          & = \pull{m} \left( \push{(\kappa_{\rho})}(\top_{C_\rho}) \sqcup \bigsqcup_{\sigma \in \leaves(T_i) \setminus \{ \rho \}} \push{(\kappa_{\sigma})} (\top_{C_{\sigma}}) \right)
          &
          \\
          & = \pull{m} \left( \push{(\kappa_{\rho})}(\top_{C_\rho}) \sqcup \bigsqcup_{\sigma \in \leaves(T_i) \setminus \{ \rho \}} \push{(m \circ \iota_{\sigma})} (\top_{C_{\sigma}}) \right)
          & \text{$\kappa_{\sigma} = m \circ \iota_{\sigma}$}
          \\
          & = \pull{m} \left( \push{(\kappa_{\rho})}(\top_{C_\rho}) \sqcup \push{m} \left( \bigsqcup_{\sigma \in \leaves(T_i) \setminus \{ \rho \}} \push{(\iota_{\sigma})} (\top_{C_{\sigma}}) \right) \right)
          & \text{$\push{m}$: left adjoint}
          \\
          & = \pull{m} \push{(\kappa_{\rho})}(\top_{C_\rho}) \sqcup \pull{m} \push{m} \left( \bigsqcup_{\sigma \in \leaves(T_i) \setminus \{ \rho \}} \push{(\iota_{\sigma})} (\top_{C_{\sigma}}) \right)
          & \text{\cref{lem:well-compatible-fib}.\itemref{item:good-fib-fibre-modular}}
          \\
          & = \push{\iota'} (\top_{C_{\rho k_0}}) \sqcup \pull{m} \push{m} \left( \bigsqcup_{\sigma \in \leaves(T_i) \setminus \{ \rho \}} \push{(\iota_{\sigma})} (\top_{C_{\sigma}}) \right)
          & \text{(\ref{eq:mQi})}
      \end{align*}
      and
      \begin{align*}
          & \pull{m} (R_{i+1})
          &
          \\
          & = \pull{m} \left(
              \left(
                  \bigsqcup_{k = 0}^{n_{\rho}} \push{(\kappa_{\rho} \circ \kappa'_{\rho k})}(\top_{C_{\rho k}})
              \right) \right. \\
          & \hspace{4em}  \left.  \sqcup
              \left(
                  \bigsqcup_{\sigma \in \leaves(T_i) \setminus \{ \rho \}} \push{(\kappa_{\sigma})} (\top_{C_{\sigma}})
              \right)
          \right)
          & \text{Definition of $R_{i+1}$}
          \\
          & = \pull{m}
          \left(
              \push{(\kappa_{\rho})}(R_{\rho})
              \sqcup
              \left(
                  \bigsqcup_{\sigma \in \leaves(T_i) \setminus \{ \rho \}} \push{(\kappa_{\sigma})} (\top_{C_{\sigma}})
              \right)
          \right)
          & \text{$\push{(\kappa_{\rho})}$: left adjoint}
          \\
          & = \pull{m}
          \left(
              \push{(\kappa_{\rho})}(R_{\rho})
              \sqcup
              \left(
                  \bigsqcup_{\sigma \in \leaves(T_i) \setminus \{ \rho \}} \push{(m \circ \iota_{\sigma})} (\top_{C_{\sigma}})
              \right)
          \right)
          & \text{$\kappa_{\sigma} = m \circ \iota_{\sigma}$}
          \\
          & = \pull{m}
          \left(
              \push{(\kappa_{\rho})}(R_{\rho})
              \sqcup
              \push{m}
              \left(
                  \bigsqcup_{\sigma \in \leaves(T_i)} \push{(\iota_{\sigma})} (\top_{C_{\sigma}})
              \right)
          \right)
          & \text{$\push{m}$: left adjoint}
          \\
          & = \pull{m}\push{(\kappa_{\rho})}(R_{\rho})
              \sqcup
              \pull{m} \push{m}
              \left(
                  \bigsqcup_{\sigma \in \leaves(T_i)} \push{(\iota_{\sigma})} (\top_{C_{\sigma}})
              \right)
          & \text{\cref{lem:well-compatible-fib}.\itemref{item:good-fib-fibre-modular}}
          \\
          & = \push{\iota'} (\top_{C_{\rho k_0}})
              \sqcup
              \pull{m} \push{m} \left( \bigsqcup_{\sigma \in \leaves(T_i) \setminus \{ \rho \}} \push{(\iota_{\sigma})} (\top_{C_{\sigma}}) \right).
          & \text{(\ref{eq:mQi1})}
      \end{align*}
      Therefore, we obtain $\pull{m}(R_i) = \pull{m}(R_{i+1})$.
  \end{inlineproof}

  We illustrate the proof of \cref{lem:partition-not-changed}. 
  See \cref{fig:example-partition-not-changed}.
  A set $C$ is the disjoint union $C_0 \cup C_1 \cup C_2$, where $C_0 = C_{00} \cup C_{01} \cup C_{02}$.
  An equivalence relation $R$ on $C$ corresponds to a partitioning $\{ C_0, C_1, C_2 \}$,
  and an equivalence relation $R'$ on $C$ corresponds to a partitioning $\{ C_{00}, C_{01}, C_{02}, C_1, C_2 \}$.
  As illustrated in \cref{fig:example-partition-not-changed},
  when restricted to $C_{00} \cup C_{1} \cup C_{2}$,
  $R$ and $R'$ coincide.
  This restriction yields the same equivalence relation,
  and \cref{lem:partition-not-changed} formalises this coincidence.
  An important observation is that,
  we cannot \emph{add} one extra to the restriction.
  For example,
  if we restrict $R$ and $R'$ to $C_{00} \cup C_{01} \cup C_1 \cup C_2$ (with both $C_{00} $ and $C_{01}$ included),
  we 
  have $ \pull{m}(R) \ne \pull{m}(R')$,
  as depicted in \cref{fig:non-example-partition-not-changed}.
  \begin{figure}[ht]
      \centering
      \begin{tikzpicture}
          \pgfmathsetmacro{\Cwidth}{2.4}
          \pgfmathsetmacro{\Cheight}{1}
          \pgfmathsetmacro{\marginV}{1.5}
          \coordinate (C'begin) at (1.7, 0) ;
          \coordinate (Cbegin)  at ($ (C'begin) + (0, \marginV) $);
          \node (C') at (0, 0.5) {$C_{00} \cup C_{1} \cup C_{2}$} ;
          \node (C'eq) at ($ (C') + (1.4, 0) $) {$=$} ;
          \node (C) at ($ (C') + (0, \marginV) $) {$C$} ;
          \node (Ceq) at ($ (C) + (1.4, 0) $) {$=$} ;
          \draw[>->] (C') -- node[left]{$m$} (C) ;
          \draw[fill=gray!10] (C'begin)
              -- ($ (C'begin) + (\Cwidth, 0) $)
              -- ($ (C'begin) + (\Cwidth, \Cheight / 2) $)
              -- ($ (C'begin) + (\Cwidth / 3, \Cheight / 2) $)
              -- ($ (C'begin) + (\Cwidth / 3, \Cheight) $)
              -- ($ (C'begin) + (0, \Cheight) $)
              -- cycle ;
          \draw[fill=gray!10] (Cbegin)
              -- ($ (Cbegin) + (\Cwidth, 0) $)
              -- ($ (Cbegin) + (\Cwidth, \Cheight) $)
              -- ($ (Cbegin) + (0, \Cheight) $)
              -- cycle ;
          \node (C00) at ($ (Cbegin) + (\Cwidth / 6,     \Cheight / 4 * 3) $) {$C_{00}$};
          \node (C01) at ($ (Cbegin) + (\Cwidth / 2,     \Cheight / 4 * 3) $) {$C_{01}$};
          \node (C02) at ($ (Cbegin) + (\Cwidth / 6 * 5, \Cheight / 4 * 3) $) {$C_{02}$};
          \node (C1)  at ($ (Cbegin) + (\Cwidth / 4, \Cheight / 4) $) {$C_1$};
          \node (C2)  at ($ (Cbegin) + (\Cwidth / 4 * 3, \Cheight / 4) $) {$C_2$};
          \draw[dashed] ($ (Cbegin) + (0, \Cheight / 2) $) -- ($ (Cbegin) + (\Cwidth, \Cheight / 2) $) ;
          \draw[dashed] ($ (Cbegin) + (\Cwidth / 3,     \Cheight / 2) $) -- ($ (Cbegin) + (\Cwidth / 3,     \Cheight) $) ;
          \draw[dashed] ($ (Cbegin) + (\Cwidth / 3 * 2, \Cheight / 2) $) -- ($ (Cbegin) + (\Cwidth / 3 * 2, \Cheight) $) ;
          \draw[dashed] ($ (Cbegin) + (\Cwidth / 2, 0) $) -- ($ (Cbegin) + (\Cwidth / 2, \Cheight / 2) $) ;
      \end{tikzpicture}
      \begin{tikzpicture}
          \pgfmathsetmacro{\Cwidth}{2.4}
          \pgfmathsetmacro{\Cheight}{1}
          \pgfmathsetmacro{\marginV}{1.5}
          \coordinate (mQbegin) at (1, 0) ;
          \coordinate (Qbegin)  at ($ (mQbegin) + (0, \marginV) $);
          \node (mQ) at (0, 0.5) {$\pull{m}(R)$} ;
          \node (mQeq) at ($ (mQ) + (0.7, 0) $) {$=$} ;
          \node (Q) at ($ (mQ) + (0, \marginV) $) {$R$} ;
          \node (Qeq) at ($ (Q) + (0.7, 0) $) {$=$} ;
          \draw[|->] (Q) -- node[left]{$\pull{m}$} (mQ) ;
          \draw[fill=gray!10] (mQbegin)
              -- ($ (mQbegin) + (\Cwidth, 0) $)
              -- ($ (mQbegin) + (\Cwidth, \Cheight / 2) $)
              -- ($ (mQbegin) + (\Cwidth / 3, \Cheight / 2) $)
              -- ($ (mQbegin) + (\Cwidth / 3, \Cheight) $)
              -- ($ (mQbegin) + (0, \Cheight) $)
              -- cycle ;
          \draw[fill=gray!10] (Qbegin)
              -- ($ (Qbegin) + (\Cwidth, 0) $)
              -- ($ (Qbegin) + (\Cwidth, \Cheight) $)
              -- ($ (Qbegin) + (0, \Cheight) $)
              -- cycle ;
          \node (QC0) at ($ (Qbegin) + (\Cwidth / 2, \Cheight / 4 * 3) $) {$C_0$};
          \node (QC1) at ($ (Qbegin) + (\Cwidth / 4, \Cheight / 4) $) {$C_1$};
          \node (QC2) at ($ (Qbegin) + (\Cwidth / 4 * 3, \Cheight / 4) $) {$C_2$};
          \draw ($ (Qbegin) + (0, \Cheight / 2) $) -- ($ (Qbegin) + (\Cwidth, \Cheight / 2) $) ;
          \draw ($ (Qbegin) + (\Cwidth / 2, 0) $) -- ($ (Qbegin) + (\Cwidth / 2, \Cheight / 2) $) ;
          \draw ($ (mQbegin) + (0, \Cheight / 2) $) -- ($ (mQbegin) + (\Cwidth / 3, \Cheight / 2) $) ;
          \draw ($ (mQbegin) + (\Cwidth / 2, 0) $) -- ($ (mQbegin) + (\Cwidth / 2, \Cheight / 2) $) ;
      \end{tikzpicture}
      \begin{tikzpicture}
          \pgfmathsetmacro{\Cwidth}{2.4}
          \pgfmathsetmacro{\Cheight}{1}
          \pgfmathsetmacro{\marginV}{1.5}
          \coordinate (mQbegin) at (1, 0) ;
          \coordinate (Qbegin)  at ($ (mQbegin) + (0, \marginV) $);
          \node (mQ) at (0, 0.5) {$\pull{m}(R')$} ;
          \node (mQeq) at ($ (mQ) + (0.7, 0) $) {$=$} ;
          \node (Q) at ($ (mQ) + (0, \marginV) $) {$R'$} ;
          \node (Qeq) at ($ (Q) + (0.7, 0) $) {$=$} ;
          \draw[|->] (Q) -- node[left]{$\pull{m}$} (mQ) ;
          \draw[fill=gray!10] (mQbegin)
              -- ($ (mQbegin) + (\Cwidth, 0) $)
              -- ($ (mQbegin) + (\Cwidth, \Cheight / 2) $)
              -- ($ (mQbegin) + (\Cwidth / 3, \Cheight / 2) $)
              -- ($ (mQbegin) + (\Cwidth / 3, \Cheight) $)
              -- ($ (mQbegin) + (0, \Cheight) $)
              -- cycle ;
          \draw[fill=gray!10] (Qbegin)
              -- ($ (Qbegin) + (\Cwidth, 0) $)
              -- ($ (Qbegin) + (\Cwidth, \Cheight) $)
              -- ($ (Qbegin) + (0, \Cheight) $)
              -- cycle ;
          \node (QC00) at ($ (Qbegin) + (\Cwidth / 6,     \Cheight / 4 * 3) $) {$C_{00}$};
          \node (QC01) at ($ (Qbegin) + (\Cwidth / 2,     \Cheight / 4 * 3) $) {$C_{01}$};
          \node (QC02) at ($ (Qbegin) + (\Cwidth / 6 * 5, \Cheight / 4 * 3) $) {$C_{02}$};
          \node (QC1) at ($ (Qbegin) + (\Cwidth / 4, \Cheight / 4) $) {$C_1$};
          \node (QC2) at ($ (Qbegin) + (\Cwidth / 4 * 3, \Cheight / 4) $) {$C_2$};
          \draw ($ (Qbegin) + (0, \Cheight / 2) $) -- ($ (Qbegin) + (\Cwidth, \Cheight / 2) $) ;
          \draw ($ (Qbegin) + (\Cwidth / 3,     \Cheight / 2) $) -- ($ (Qbegin) + (\Cwidth / 3,     \Cheight) $) ;
          \draw ($ (Qbegin) + (\Cwidth / 3 * 2, \Cheight / 2) $) -- ($ (Qbegin) + (\Cwidth / 3 * 2, \Cheight) $) ;
          \draw ($ (Qbegin) + (\Cwidth / 2, 0) $) -- ($ (Qbegin) + (\Cwidth / 2, \Cheight / 2) $) ;
          \draw ($ (mQbegin) + (0, \Cheight / 2) $) -- ($ (mQbegin) + (\Cwidth / 3, \Cheight / 2) $) ;
          \draw ($ (mQbegin) + (\Cwidth / 2, 0) $) -- ($ (mQbegin) + (\Cwidth / 2, \Cheight / 2) $) ;
      \end{tikzpicture}
      \caption{An example situation of \cref{lem:partition-not-changed} in $\eqrel \to \sets$: $\pull{m}(R) = \pull{m}(R')$.}
      \label{fig:example-partition-not-changed}
  \end{figure}
  \begin{figure}[ht]
      \centering
      \begin{tikzpicture}[baseline=0]
          \pgfmathsetmacro{\Cwidth}{2.4}
          \pgfmathsetmacro{\Cheight}{1}
          \pgfmathsetmacro{\marginV}{1.5}
          \coordinate (C'begin) at (1.7, 0) ;
          \coordinate (Cbegin)  at ($ (C'begin) + (0, \marginV) $);
          \node (C') at (0, 0.5) {$C_{00} \cup C_{01}$} ;
          \node (C'additional) at (0.2, 0) {${} \cup C_{1} \cup C_{2}$} ;
          \node (C'eq) at ($ (C') + (1.4, 0) $) {$=$} ;
          \node (C) at ($ (C') + (0, \marginV) $) {$C$} ;
          \node (Ceq) at ($ (C) + (1.4, 0) $) {$=$} ;
          \draw[>->] (C') -- node[left]{$m$} (C) ;
          \draw[fill=gray!10] (C'begin)
              -- ($ (C'begin) + (\Cwidth, 0) $)
              -- ($ (C'begin) + (\Cwidth, \Cheight / 2) $)
              -- ($ (C'begin) + (\Cwidth / 3 * 2, \Cheight / 2) $)
              -- ($ (C'begin) + (\Cwidth / 3 * 2, \Cheight) $)
              -- ($ (C'begin) + (0, \Cheight) $)
              -- cycle ;
          \draw[fill=gray!10] (Cbegin)
              -- ($ (Cbegin) + (\Cwidth, 0) $)
              -- ($ (Cbegin) + (\Cwidth, \Cheight) $)
              -- ($ (Cbegin) + (0, \Cheight) $)
              -- cycle ;
          \node (C00) at ($ (Cbegin) + (\Cwidth / 6,     \Cheight / 4 * 3) $) {$C_{00}$};
          \node (C01) at ($ (Cbegin) + (\Cwidth / 2,     \Cheight / 4 * 3) $) {$C_{01}$};
          \node (C02) at ($ (Cbegin) + (\Cwidth / 6 * 5, \Cheight / 4 * 3) $) {$C_{02}$};
          \node (C1)  at ($ (Cbegin) + (\Cwidth / 4, \Cheight / 4) $) {$C_1$};
          \node (C2)  at ($ (Cbegin) + (\Cwidth / 4 * 3, \Cheight / 4) $) {$C_2$};
          \draw[dashed] ($ (Cbegin) + (0, \Cheight / 2) $) -- ($ (Cbegin) + (\Cwidth, \Cheight / 2) $) ;
          \draw[dashed] ($ (Cbegin) + (\Cwidth / 3,     \Cheight / 2) $) -- ($ (Cbegin) + (\Cwidth / 3,     \Cheight) $) ;
          \draw[dashed] ($ (Cbegin) + (\Cwidth / 3 * 2, \Cheight / 2) $) -- ($ (Cbegin) + (\Cwidth / 3 * 2, \Cheight) $) ;
          \draw[dashed] ($ (Cbegin) + (\Cwidth / 2, 0) $) -- ($ (Cbegin) + (\Cwidth / 2, \Cheight / 2) $) ;
      \end{tikzpicture}
      \begin{tikzpicture}[baseline=0]
          \pgfmathsetmacro{\Cwidth}{2.4}
          \pgfmathsetmacro{\Cheight}{1}
          \pgfmathsetmacro{\marginV}{1.5}
          \coordinate (mQbegin) at (1, 0) ;
          \coordinate (Qbegin)  at ($ (mQbegin) + (0, \marginV) $);
          \node (mQ) at (0, 0.5) {$\pull{m}(R)$} ;
          \node (mQeq) at ($ (mQ) + (0.7, 0) $) {$=$} ;
          \node (Q) at ($ (mQ) + (0, \marginV) $) {$R$} ;
          \node (Qeq) at ($ (Q) + (0.7, 0) $) {$=$} ;
          \draw[|->] (Q) -- node[left]{$\pull{m}$} (mQ) ;
          \draw[fill=gray!10] (mQbegin)
              -- ($ (mQbegin) + (\Cwidth, 0) $)
              -- ($ (mQbegin) + (\Cwidth, \Cheight / 2) $)
              -- ($ (mQbegin) + (\Cwidth / 3 * 2, \Cheight / 2) $)
              -- ($ (mQbegin) + (\Cwidth / 3 * 2, \Cheight) $)
              -- ($ (mQbegin) + (0, \Cheight) $)
              -- cycle ;
          \draw[fill=gray!10] (Qbegin)
              -- ($ (Qbegin) + (\Cwidth, 0) $)
              -- ($ (Qbegin) + (\Cwidth, \Cheight) $)
              -- ($ (Qbegin) + (0, \Cheight) $)
              -- cycle ;
          \node (QC0) at ($ (Qbegin) + (\Cwidth / 2, \Cheight / 4 * 3) $) {$C_0$};
          \node (QC1) at ($ (Qbegin) + (\Cwidth / 4, \Cheight / 4) $) {$C_1$};
          \node (QC2) at ($ (Qbegin) + (\Cwidth / 4 * 3, \Cheight / 4) $) {$C_2$};
          \draw ($ (Qbegin) + (0, \Cheight / 2) $) -- ($ (Qbegin) + (\Cwidth, \Cheight / 2) $) ;
          \draw ($ (Qbegin) + (\Cwidth / 2, 0) $) -- ($ (Qbegin) + (\Cwidth / 2, \Cheight / 2) $) ;
          \draw ($ (mQbegin) + (0, \Cheight / 2) $) -- ($ (mQbegin) + (\Cwidth / 3 * 2, \Cheight / 2) $) ;
          \draw ($ (mQbegin) + (\Cwidth / 2, 0) $) -- ($ (mQbegin) + (\Cwidth / 2, \Cheight / 2) $) ;
      \end{tikzpicture}
      \begin{tikzpicture}[baseline=0]
          \pgfmathsetmacro{\Cwidth}{2.4}
          \pgfmathsetmacro{\Cheight}{1}
          \pgfmathsetmacro{\marginV}{1.5}
          \coordinate (mQbegin) at (1, 0) ;
          \coordinate (Qbegin)  at ($ (mQbegin) + (0, \marginV) $);
          \node (mQ) at (0, 0.5) {$\pull{m}(R')$} ;
          \node (mQeq) at ($ (mQ) + (0.7, 0) $) {$=$} ;
          \node (Q) at ($ (mQ) + (0, \marginV) $) {$R'$} ;
          \node (Qeq) at ($ (Q) + (0.7, 0) $) {$=$} ;
          \draw[|->] (Q) -- node[left]{$\pull{m}$} (mQ) ;
          \draw[fill=gray!10] (mQbegin)
              -- ($ (mQbegin) + (\Cwidth, 0) $)
              -- ($ (mQbegin) + (\Cwidth, \Cheight / 2) $)
              -- ($ (mQbegin) + (\Cwidth / 3 * 2, \Cheight / 2) $)
              -- ($ (mQbegin) + (\Cwidth / 3 * 2, \Cheight) $)
              -- ($ (mQbegin) + (0, \Cheight) $)
              -- cycle ;
          \draw[fill=gray!10] (Qbegin)
              -- ($ (Qbegin) + (\Cwidth, 0) $)
              -- ($ (Qbegin) + (\Cwidth, \Cheight) $)
              -- ($ (Qbegin) + (0, \Cheight) $)
              -- cycle ;
          \node (QC00) at ($ (Qbegin) + (\Cwidth / 6,     \Cheight / 4 * 3) $) {$C_{00}$};
          \node (QC01) at ($ (Qbegin) + (\Cwidth / 2,     \Cheight / 4 * 3) $) {$C_{01}$};
          \node (QC02) at ($ (Qbegin) + (\Cwidth / 6 * 5, \Cheight / 4 * 3) $) {$C_{02}$};
          \node (QC1) at ($ (Qbegin) + (\Cwidth / 4, \Cheight / 4) $) {$C_1$};
          \node (QC2) at ($ (Qbegin) + (\Cwidth / 4 * 3, \Cheight / 4) $) {$C_2$};
          \draw ($ (Qbegin) + (0, \Cheight / 2) $) -- ($ (Qbegin) + (\Cwidth, \Cheight / 2) $) ;
          \draw ($ (Qbegin) + (\Cwidth / 3,     \Cheight / 2) $) -- ($ (Qbegin) + (\Cwidth / 3,     \Cheight) $) ;
          \draw ($ (Qbegin) + (\Cwidth / 3 * 2, \Cheight / 2) $) -- ($ (Qbegin) + (\Cwidth / 3 * 2, \Cheight) $) ;
          \draw ($ (Qbegin) + (\Cwidth / 2, 0) $) -- ($ (Qbegin) + (\Cwidth / 2, \Cheight / 2) $) ;
          \draw ($ (mQbegin) + (0, \Cheight / 2) $) -- ($ (mQbegin) + (\Cwidth / 3 * 2, \Cheight / 2) $) ;
          \draw ($ (mQbegin) + (\Cwidth / 2, 0) $) -- ($ (mQbegin) + (\Cwidth / 2, \Cheight / 2) $) ;
          \draw ($ (mQbegin) + (\Cwidth / 3, \Cheight / 2) $) -- ($ (mQbegin) + (\Cwidth / 3, \Cheight) $) ;
      \end{tikzpicture}
      \caption{
    We cannot add more than one child to the domain of $m$ in \cref{lem:partition-not-changed}.
      }
      \label{fig:non-example-partition-not-changed}
  \end{figure}
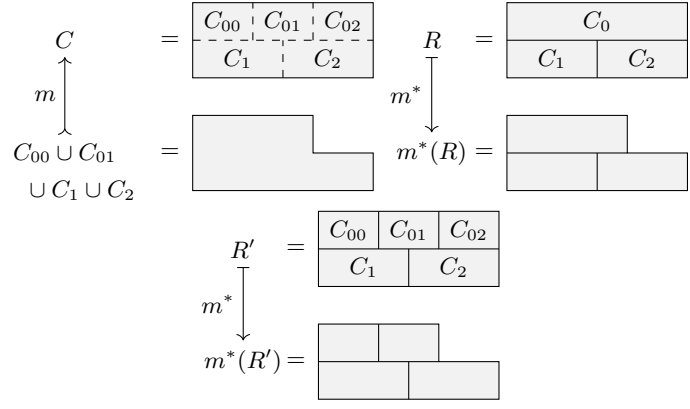

  We go back to the proof of \cref{prop:loop-invariant}.
  The claim \itemref{item:keep-partitioning} follows from \cref{lem:part-part} and the induction on $i$.
  The claim \itemref{item:gfp-R} follows from
  the assumption that $p$ admits partitioning (\cref{assum}.\itemref{assum:monotone-partitioning})
  and the induction on $i$.
  We prove \itemref{item:clean-leaf-top} by the induction on $i$.

  (\textbf{Base case}).
  We have $C^{\clean, 0}_{\epsilon} = 0$.
  Hence, from \cref{assum}.\itemref{assum:fibre-0},
  we have 
  \begin{equation*}
    \pull{({\kappa^{\clean,0}_{\epsilon}})}\pull{\kappa_{\epsilon}}\pull{c}\lift{F}(R_0) = \top_{C^{\clean, 0}_{\epsilon}}.
  \end{equation*}
  
  (\textbf{Induction step}).
  We assume that \itemref{item:clean-leaf-top} holds for $i$,
  and need to show \itemref{item:clean-leaf-top} for $i+1$.
  For each $\sigma \in \leaves(T_{i+1})$,
  $C^{\clean, i+1}_{\sigma}$ is defined by the following pullback:
  \begin{equation*}
      \begin{tikzcd}
          C^{\clean, i+1}_{\sigma}
          \ar[dd, >->]
          \ar[r, "\kappa"]
          \ar[rd, phantom, "\usebox\pullback", very near start]
          &
          C^{\clean, i}_{\sigma}
          \ar[d, >->, "\kappa^{\clean, i}_{\sigma}"]
          \\
          &
          C_{\sigma}
          \ar[d, >->, "\kappa_{\sigma}"]
          \\
          B
          \ar[r, "l"]
          \ar[d, "h"']
          \ar[rd, phantom, "\usebox\pullback", very near start]
          &
          C
          \ar[d, "c"]
          \\
          F\left(  C_{\rho k_0} \cup \left( \bigcup_{\sigma \in \leaves(T) \setminus \{ \rho \} } C_\sigma \right) \right)
          \ar[r, "Fm"']
          &
          FC \mathrlap{\quad.}
      \end{tikzcd}
  \end{equation*}
  Applying \cref{lem:successor-and-partition} to the above diagram and the result of \cref{lem:partition-not-changed},
  we have
  \begin{equation}
      \pull{(\kappa^{\clean, i+1}_{\sigma})} \pull{\kappa_{\sigma}} \pull{c} \lift{F} R_{i+1}
      = \pull{(\kappa^{\clean, i+1}_{\sigma})} \pull{\kappa_{\sigma}} \pull{c} \lift{F} R_{i}.
  \end{equation}
  If $\sigma \in T_i$, the right-hand side is calculated as:
  \begin{align*}
      \pull{(\kappa^{\clean, i+1}_{\sigma})} \pull{\kappa_{\sigma}} \pull{c} \lift{F} R_{i}
      & = \pull{\kappa} \pull{(\kappa^{\clean, i}_{\sigma})} \pull{\kappa_{\sigma}} \pull{c} \lift{F} R_{i}
      &
      \\
      & = \pull{\kappa} (\top_{C^{\clean, i}_{\sigma}})
      & \text{Induction hypothesis}
      \\
      & = \top_{C^{\clean, i+1}_{\sigma}}.
      & \text{$\push{\kappa} \dashv \pull{\kappa}$ and $\top_{C^{\clean, i}_{\sigma}}$ is a limit}
  \end{align*}
  Otherwise, we have $\sigma = \rho k$ for some $k \in \{ 0, \dots , n_{\rho} \}$
  where $\rho \in \leaves(T_i)$ is the chosen leaf at the \textbf{Partitioning} step.
  Then, we have
  \begin{align*}
      & \pull{(\kappa^{\clean, i+1}_{\rho k})} \pull{\kappa_{\rho k}} \pull{c} \lift{F} R_{i}
      & 
      \\
      & = \pull{(\kappa^{\clean, i+1}_{\rho k})} \pull{\kappa_{\rho k, k}} \pull{\kappa_{\rho}} \pull{c} \lift{F} R_{i}
      & \kappa_{\rho k} = \kappa_{\rho} \circ \kappa_{\rho k, k}
      \\
      & = \pull{(\kappa^{\clean, i+1}_{\rho k})} (\top_{C_{\rho k}})
      & \text{\cref{def:R-partitioning}.\itemref{item:R-part-cond-1}}
      \\
      & = \top_{C^{\clean, i+1}_{\rho k}}.
      & \text{$\push{(\kappa^{\clean, i+1}_{\rho k})} \dashv \pull{(\kappa^{\clean, i+1}_{\rho k})}$ and $\top_{C_{\rho k}}$ is a limit}
  \end{align*}
  Hence, we have
  $\pull{(\kappa^{\clean, i+1}_{\sigma})} \pull{\kappa_{\sigma}} \pull{c} \lift{F} R_{i+1} = \top_{C^{\clean, i+1}_{\sigma}}$
  for each $\sigma \in \leaves(T_{i+1})$.
\end{proof}

\begin{mypropositionrep}[termination]\label{prop:termination}
  \cref{algo:fibrational-block-specified} terminates for any input.
\end{mypropositionrep}
The key observation for the proof of termination is that in each iteration of the main loop either the partition is refined, or it is not but the number of dirty leaves decreases.
\begin{proof}
  To prove termination,
  we use the lexicographical order $\trianglelefteq$
  on $\cate_{C} \times \nat$, where the order of $\nat$ is the usual order $\le$.
  The order $\le$ is well-founded,
  and $\sqsubseteq$ is well-founded by \cref{assum}.\itemref{assum:well-founded}.
  Thus, $\trianglelefteq$ is also well-founded.
  Let $d_i = |\{ \sigma \in \leaves(T_i) \mid C^{\clean, i}_{\sigma} \ne C_{\sigma} \}|$.

  We show that
  if $C^{\clean, i}_{\rho} \ne C^{i}_{\rho}$ holds for some $\rho \in \leaves(T)$,
  then we have \[ (R_{i + 1}, d_{i + 1}) \triangleleft (R_i , d_i). \]
  Suppose that, at the line \ref{line:choose-a-leaf} of the $i$-th iteration of the main loop,
  a leaf $\rho \in \leaves(T_i)$ such that $C^{\clean, i}_{\rho} \ne C^{i}_{\rho}$ is chosen.
  We have two cases.

  \textbf{Case} $\pull{(c \circ \kappa_{\rho})} (\lift{F} R_i) \sqsubset \top_{C_\rho}$.
  In this case, we have $R_i \sqsubset R_{i + 1}$ by \cref{assum}.\itemref{assum:inj-on-obj}.
  Thus, $(R_{i + 1}, d_{i + 1}) \triangleleft (R_i , d_i)$ holds.

  \textbf{Case} $\pull{(c \circ \kappa_{\rho})} (\lift{F} R_i) = \top_{C_\rho}$.
  In this case,
  the line \ref{line:mark-all-states-clean} is executed,
  and the number of $\sigma \in T$ with $C^{\clean}_{\sigma} \ne C_{\sigma}$ reduces: $d_{i + 1} < d_i$.

  By the above argument shows that,
  after finite number of iteration, we have $C^{\clean}_{\rho} = C_{\rho}$ for all $\rho \in \leaves(T)$.
  This means that \cref{algo:fibrational-block-specified} terminates.
\end{proof}

\begin{mytheoremrep}[correctness]\label{thm:correctness}
  \cref{algo:fibrational-block-specified} terminates for any input and returns a $\nu (\pull{c} \lift{F})$-partitioning.
\end{mytheoremrep}
\begin{proof}
  Termination is ensured by \cref{prop:termination}.
  Let $R$, $T$ and $\{ C_{\sigma} \monoto C \}_{\sigma \in \leaves(T)}$ be
  as defined in the last iteration of the main loop.
  Our goal is to show that $\{ C_{\sigma} \monoto C \}_{\sigma \in \leaves(T)}$ is a $\nu (\pull{c} \lift{F})$-partitioning of $C$.

  We prove $R = \pull{c}\lift{F}(R)$.
  By \cref{prop:loop-invariant},
  $\{ C_{\sigma} \monoto C \}_{\sigma \in \leaves(T)}$ is an $R$-partitioning.
  We also have $\pull{\kappa_{\sigma}}(\pull{c}\lift{F}(R)) = \top_{C_{\sigma}}$ for each $\sigma \in \leaves(T)$.
  Thus, we have
  \begin{align*}
      R
      & = \bigsqcup_{\sigma \in \leaves(T)} \push{(\kappa_{\sigma})}(\top_{C_{\sigma}})
      & \text{the definition of $R$}
      \\
      & = \bigsqcup_{\sigma \in \leaves(T)} \push{(\kappa_{\sigma})}(\pull{\kappa_{\sigma}}(\pull{c}\lift{F}(R)))
      & \text{\cref{def:R-partitioning}.\itemref{item:R-part-cond-2}}
      \\
      & = \pull{c}\lift{F}(R).
      & \text{\cref{assum}.\itemref{assum:monotone-partitioning}}
  \end{align*}
  From \cref{prop:loop-invariant}.\itemref{item:gfp-R},
  we have $\nu(\pull{c}\lift{F}) \sqsubseteq R$.
  Hence, we have $R = \nu(\pull{c}\lift{F})$.
\end{proof}

The explicit correspondence between $\PRHopcroft$ and \cref{sec:Hopcroft-trick} (\cref{table:Hopcroft-ineq-and-blockspecified-algorithm}) allows us to directly use Hopcroft's inequality.
The following result, while it does not give a complexity bound for $\PRHopcroft$ itself, plays a central role in the amortised analysis of its concrete instances in~\cref{sec:complexity-analysis}.

\begin{myproposition}\label{prop:time-markdirty}
  If each call of $\textsc{MarkDirty}$ in \cref{algo:fibrational-block-specified} takes
  \[ \order\left(K \sum_{k = \{0, \dots, n_{\rho}\} \setminus \{ k_0\}} w(C_{\rho k})\right)\] time for some $K$,
  the total time taken by the repeated calls of $\textsc{MarkDirty}$ is $\order(K w(C) \log w(C))$.
\end{myproposition}

\subsection{Concrete Yet Functor-Generic Algorithms $\PRHopcroftEqRel{\wCard}$, $\PRHopcroftEqRel{\wPred}$, $\PRHopcroftEqRel{\wReach}$}
\label{sec:complexity-analysis}
We instantiate the fibrational algorithm $\PRHopcroft_{(F, \lift{F}), w}$ with $\eqrel \to \sets$ as a base fibration.
In this situation, the functor $F$ is an endofunctor on $\sets$ and $\lift{F}$ is an endofunctor on $\eqrel$ which is a fibred lifting of $F$.
This instantiation also enables a semantically equivalent reformulation of \textsc{MarkDirty}---its ``implementation'' is now ``predecessor-centric'' rather than  ``successor-centric''---and this aids more refined complexity analysis. 

For a weight function $w$ (a parameter of $\PRHopcroft$),
we introduce three examples $\wCard, \wPred, \wReach$, leading to three functor-generic algorithms $\PRHopcroftER_{(F, \lift{F}), \wCard}$, $\PRHopcroftER_{(F, \lift{F}), \wPred}$ and $\PRHopcroftER_{(F, \lift{F}), \wReach}$.

\begin{mydefinition}[$\PRHopcroftEqRel{w}$] \label{def:eqrel-sets-algo}
  Let $\sets \xrightarrow{F} \sets$ and $\eqrel \xrightarrow{\lift{F}} \eqrel$ be functors,
  $C \xrightarrow{c} FC$ be a coalgebra,
  and $w \colon \powset(C) \to \nat$ be a function
  (which amounts to $w \colon  \obj(\sub(\catc)_C) \to \nat$ in \cref{subsec:PRHopcroft}), all satisfying
  \cref{assum} ($C$ must be finite, in particular).
  The algorithm $\PRHopcroftEqRel{w}$ is shown in 
  \cref{algo:eqrel-sets-optimized};
  it computes a $\nu(\pull{c}\lift{F})$-partitioning of $C$.
\end{mydefinition}

\begin{algorithm}[t]
  \caption{
      The algorithm $\PRHopcroftEqRel{w}$,
      obtained as an instance of $\PRHopcroft$ (\cref{algo:fibrational-block-specified}) where $p$ is $\eqrel\to\sets$,
      with a semantically equivalent formulation of \textsc{MarkDirty} (successor-centric in $\PRHopcroft$; predecessor-centric here in  $\PRHopcroftEqRel{w}$).
      }
  \label{algo:eqrel-sets-optimized}
  \begin{algorithmic}[1]
  \Require A coalgebra $c \colon C \to FC$ in $\sets$.
  \Ensure A mono-sink $\{ \kappa_i \colon C_i \monoto C \}_{i \in I}$ for some $I$.
  \Statex
  \Statex (the same as Line~1--13 of \cref{algo:fibrational-block-specified})
      \Statex
          \setcounter{ALG@line}{13}
          \Procedure{MarkDirty}{$\rho, k_0$}
              \For{$k \in \{0, \dots, n_{\rho}\}$}
                  $C^{\clean}_{\rho k} := C_{\rho k}$
              \EndFor
              \For{$k \in \{0, \dots, n_{\rho}\} \setminus \{ k_0\}$ and $y \in C_{\rho k}$}\label{line:optimised-markdirty-predecessors}
                  \For{$x$: predecessor of $y$}\label{line:optimised-markdirty-one-predecessor}
                      \State\label{line:lookup} Find $\tau \in \leaves(J \cup \{ \rho 0, \dots, \rho n_{\rho}\})$ such that $x \in C_{\tau}$
                      \State\label{line:optimised-mark-a-state-dirty}If such $\tau$ exists, then $C^{\clean}_{\tau} := C^{\clean}_{\tau} \setminus \{ x \}$
               \Comment{mark $x$ as dirty}
                  \EndFor
              \EndFor
          \EndProcedure
      \end{algorithmic}
\end{algorithm}
Line~14--19 of \cref{algo:eqrel-sets-optimized} uses this categorical notion of predecessor (Line~17), which is in \cref{def:pred}. Its equivalence to the original definition (Line~14--19 of \cref{algo:fibrational-block-specified}) is easy; so $\PRHopcroftEqRel{w}$ is correct by \cref{thm:correctness}. The successor-centric description is more convenient in the correctness proof (\cref{thm:correctness}), while the predecessor-centric one is advantageous for complexity analysis. 
\begin{mydefinition}[predecessor \cite{JacobsWissmann23}]\label{def:pred}
  Let $c \colon C \to FC$ be a coalgebra in $\sets$.
  For $x, y \in C$,
  we say $x$ is a \emph{predecessor} of $y$ if
  $x \not \in B$, where $B$ is a subset of $C$ defined by the following pullback:
  \begin{equation*}
    \begin{tikzcd}[row sep=small]
        B
        \ar[d]
        \ar[r, >->]
        \ar[rd, phantom, "\usebox\pullback", very near start]
        &
        C
        \ar[d, "c"]
        \\
        F(C \setminus \{ y \})
        \ar[r, >->, "F(i_y)"']
        &
        FC
    \end{tikzcd}.
  \end{equation*}
  Here $i_{y}$ is the canonical injection.
\end{mydefinition}

For 
$w$ as a parameter of $\PRHopcroftEqRel{w}$, we introduce three functions.
\begin{mydefinition}[$\wCard,\wPred, \wReach$]\label{def:concreteWeightFunc}
  We define $\wCard,\wPred, \wReach \colon \powset(C) \to \nat$ as follows: 
  $\wCard(A)=|A|$, 
  $\wPred(A)=\sum_{x \in A} |\{ y \in C \mid \text{$y$ is a predecessor of $x$} \}|$,  and
  $\wReach(A)=| A \cap  C' |$,
  where $C' = \{ x \in C \mid \text{$x$ is a successor of some $y \in C$} \}$.
  The weight functions $\wCard$, $\wPred$ and $\wReach$ are called the \emph{cardinality}, \emph{predecessor} and \emph{reachability} weights, respectively.
\end{mydefinition}

The cardinality weight is the most commonly used one in various partition refinement algorithms, including~\cite{JacobsWissmann23}.
The latter two have been used in~\cite{Hopcroft71, Gries73, Knuutila01} for DFA partition refinement;
we use them for the first time in categorical partition refinement algorithms.

\begin{table}[t]
  \centering
  \footnotesize
  \rowcolors{2}{gray!10}{gray!0}
  \caption{Examples of partition refinement algorithms induced by $\PRHopcroftEqRel{(F, \lift{F}), w}$}
  \label{table:examples-of-pr}
  \begin{tabular}{llll}
      \toprule
      Functor $F(X)$ & Weight function & System & Algorithm: a variation of \\
      \midrule
      $2 \times X^A$ & $\wCard(A)$, $\wReach(A)$ & DFA & \begin{tabular}{l} Hopcroft's algorithm \\ \cite{Hopcroft71,Gries73,Knuutila01}\end{tabular} \\
      $\powset(A \times X)$ & $\wCard(A)$ & LTS  & \cite{Valmari09}\\
      $\distr(X)$ & $\wCard(A)$ & Markov chain & \cite{ValmariFranceschinis10} \\
      $\powset(\distr(X))$ & $\wCard(A)$ & Markov decision process & \cite{Groote+18} \\
      \bottomrule
  \end{tabular}
\end{table}

Our algorithm $\PRHopcroftEqRel{(F, \lift{w}), w}$ induces concrete partition refinement algorithms for various systems as shown in \cref{table:examples-of-pr}.

The effect of different weight functions is illustrated in \cref{appendix:illustRun}.

The predecessor-centric \textsc{MarkDirty} and concrete choices of $w$ allow the following fine-grained complexity analysis.
\begin{mytheorem}[complexity
of $\PRHopcroftEqRel{\wCard}, \PRHopcroftEqRel{\wPred}, \PRHopcroftEqRel{\wReach}$
]\label{thm:complexityConcrete}
  \begin{enumerate}
   \item  \label{item:complexityWCard}   Let $M = \max_{y \in C} |\{ x \in C \mid \text{$x$ is a predecessor of $y$}\}|$, the ``in-degree'' of $c\colon C\to FC$; 
      suppose  it takes $\order(f)$ time to compute $c(x) \in FC$ for each $x \in C$.
      The time complexity of $\PRHopcroftEqRel{\wCard}$ is $\order(f M |C| \log |C|)$.
   \item \label{item:complexityWPred} Let $m = \wPred (C) = \sum_{y \in C} | \{ x \in C \mid \text{$x$ is a predecessor of $y$}\} |$.
      The  time complexity of 
      $\PRHopcroftEqRel{\wPred}$ is $\order(f m \log m)$.
      Since $m \le |C|^2$,
      it is also bounded by $\order(f m \log |C|)$.
   \item  The time complexity of
     $\PRHopcroftEqRel{\wReach}$ is
   $\order(f M |C'| \log |C'|)$, where $f$ and $M$ are from above and $C'$ is from \cref{def:concreteWeightFunc}. 
  \end{enumerate}
\end{mytheorem}
We note that the complexity bound given in~\cite{JacobsWissmann23} is $\order(f m \log |C|)$ which is the same as \itemref{item:complexityWPred}: their algorithm is essentially $\PRHopcroftEqRel{\wCard}$, but their more fine-grained, element-wise analysis derived the aforementioned bound.

We sketch the proof of \itemref{item:complexityWCard} for illustration.
It uses Hopcroft's inequality in its amortised analysis.

\begin{proofsketch} (of \cref{thm:complexityConcrete}.\itemref{item:complexityWCard})
We can check that \cref{algo:eqrel-sets-optimized} (with $w=\wCard$) satisfies the premise of \cref{prop:time-markdirty}.
By implementing this algorithm properly (preparing a table for $x$ and $\tau$ in Line~\ref{line:lookup} of \cref{algo:eqrel-sets-optimized}, as done in~\cite{JacobsWissmann23}),
it takes $\order(1)$ time to execute
each iteration of the loop at  Line~\ref{line:optimised-markdirty-one-predecessor}.
Thus, the loop at  Line \ref{line:optimised-markdirty-predecessors}
takes $\order(M |C_{\rho k}|)$ time for each $k$.
Therefore 
the time taken for each call of \textsc{MarkDirty} is
$\order(M \sum_{k \in \{0, \dots, n_{\rho} \} \setminus \{ k_0 \} } |C_{\rho k}| )$.
By \cref{prop:time-markdirty}, the time taken for the repeated calls of \textsc{MarkDirty} in total is $\order(M |C| \log |C|)$.

The complexity of the other parts of the algorithm is also bounded.
We write $C^{\dirty}_{\sigma}$ for $C_{\sigma} \setminus C^{\clean}_{\sigma}$.
The computation of $R_{\rho}$ (Line~\ref{line:def-rel}--\ref{line:Rrho} of \cref{algo:fibrational-block-specified})
takes $\order(f |C^{\dirty}_{\rho}|)$, and
the computation of $R_{\rho}$-partitioning (Line \ref{line:partitioning-last-line} of \cref{algo:fibrational-block-specified}) takes $\order(|C^{\dirty}_\rho|)$,
using appropriate data structures (see \cite{Knuutila01}).
Hence, it takes $\order(f |C^{\dirty}_{\rho}|)$ for each iteration of the main loop except for $\textsc{MarkDirty}$. 

Therefore, the total time for \cref{algo:eqrel-sets-optimized}
except for \textsc{MarkDirty} (let us write $T_{\setminus\textsc{MarkDirty}}$) is
$\order(\sum_{\text{$\rho$ for each iteration}}f |C^{\dirty}_{\rho}|)$.
We use amortised analysis to bound this sum.
Specifically, it is easy to see that the sum $\sum_{\text{$\rho$}}f |C^{\dirty}_{\rho}|$ is
bounded by the number of times that states are marked as dirty, multiplied by $f$.
Throughout the algorithm, the number of times that states are marked as dirty (at Line \ref{line:optimised-mark-a-state-dirty} of \cref{algo:eqrel-sets-optimized})
is at most the time consumed by $\textsc{MarkDirty}$, which is $\order(M |C| \log |C|)$.
Therefore, $T_{\setminus\textsc{MarkDirty}}$ is  $\order(f M |C| \log |C|)$; so is the total time.
\end{proofsketch}

\begin{toappendix}
  \section{Illustration of Runs of \cref{algo:eqrel-sets-optimized}}
  \label{appendix:illustRun}
  
  We illustrate \cref{algo:eqrel-sets-optimized} for non-deterministic automata.
  
  \begin{myexample}
      Let $\Sigma = \{ a, b \}$, $C = \{ x, y, z, w, v \}$, $\nondet_{\Sigma}$ be the functor from \cref{example:F-coalg},
      and $\lift{\nondet_{\Sigma}} = \rellift{\nondet_{\Sigma}}$ (\cref{def:relation-lifting}).
      \begin{figure}
          \footnotesize
          \centering
          (4 steps) $\cdots$
          \begin{tikzpicture}[scale=0.9, x=4mm, segment length=2mm]\draw[->, decorate, decoration={snake,amplitude=1pt}, line width=1.5pt] (0,0) to node[above=2mm]{$\wCard$} (-1,0) ;\end{tikzpicture}
          \begin{minipage}{2.3cm}
              \centering
              \partitioningexample{
              \begin{pgfonlayer}{background}
                  \fill[lipicsYellow] \convexpath{x,y,v}{1cm} node[above of=x, yshift=-3.5cm, text=black]{$C_{01}$};
                  \fill[lipicsYellow] (z) circle[radius=1] node[xshift=1.5cm,yshift=-1cm,text=black] {$C_1$};
                  \fill[lipicsYellow] (w) circle[radius=1] node[xshift=0cm,yshift=-1.5cm,text=black] {$C_{00}$};
              \end{pgfonlayer}
              }
              {dirty} 
              {dirty} 
              {dirty} 
              {dirty} 
              {clean} 
          \end{minipage}
          \begin{tikzpicture}[scale=0.9, x=4mm, segment length=2mm]\draw[->, decorate, decoration={snake,amplitude=1pt}, line width=1.5pt] (0,0) to node[above=2mm]{$\wCard$} (-1,0) ;\end{tikzpicture}
          \begin{minipage}{2.4cm}
              \centering
              \partitioningexample{
              \begin{pgfonlayer}{background}
                  \fill[lipicsYellow] node[above of=x, yshift=-3.5cm, text=black]{$C_0$}
                      \hobbyconvexpath{x,y,w,v}{1cm};
                  \fill[lipicsYellow] (z) circle[radius=1] node[xshift=1.5cm, yshift=-1cm, text=black] {$C_1$};
              \end{pgfonlayer}
              }
              {dirty} 
              {dirty} 
              {clean} 
              {dirty} 
              {dirty} 
          \end{minipage}
          \begin{tikzpicture}[scale=0.9, x=4mm, segment length=2mm]\draw[->, decorate, decoration={snake,amplitude=1pt}, line width=1.5pt] (0,0) to node[above=2mm]{$\wPred$} (1,0) ;\end{tikzpicture}
          \begin{minipage}{2.3cm}
              \centering
              \partitioningexample{
              \begin{pgfonlayer}{background}
                  \fill[lipicsYellow] \convexpath{x,y,v}{1cm} node[above of=x, yshift=-3.5cm, text=black] {$C_{01}$};
                  \fill[lipicsYellow] (z) circle[radius=1] node[xshift=1.5cm,yshift=-1cm,text=black] {$C_1$};
                  \fill[lipicsYellow] (w) circle[radius=1] node[xshift=0cm,yshift=-1.5cm,text=black] {$C_{00}$};
              \end{pgfonlayer}
              }
              {dirty} 
              {dirty} 
              {clean} 
              {clean} 
              {dirty} 
          \end{minipage}
          \begin{tikzpicture}[scale=0.9, x=4mm, segment length=2mm]\draw[->, decorate, decoration={snake,amplitude=1pt}, line width=1.5pt] (0,0) to node[above=2mm]{$\wPred$} (1,0) ;\end{tikzpicture}
          $\cdots$ (2 steps)
          \caption{
              The snapshots at the end of first and second iterations of
              $\PRHopcroftEqRel{(\nondet_{\Sigma}, \lift{\nondet_{\Sigma}}), \wCard}$
              and $\PRHopcroftEqRel{(\nondet_{\Sigma}, \lift{\nondet_{\Sigma}}), \wPred}$ for $c$.
              Yellow areas depict partitions which are refined as the main loop repeats.
              Clean states are white and dirty states are black.}\label{fig:execution-example}
      \end{figure}
      We define a coalgebra $c \colon C \to \nondet_{\Sigma}C$ as shown in \cref{fig:execution-example}.
      We compare the executions $\PRHopcroftEqRel{(\nondet_\Sigma, \lift{\nondet_{\Sigma}}), \wCard}$ and $\PRHopcroftEqRel{(\nondet_\Sigma, \lift{\nondet_{\Sigma}}), \wPred}$.
      At the both initialisations (Line \ref{line:initialisation}),
      we have a tree whose sole leaf is equal to $C$, which represents the initial partition, and every state in $C$ marked as dirty.
      
      In the first iteration, both $\PRHopcroftEqRel{(\nondet_\Sigma, \lift{\nondet_{\Sigma}}), \wCard}$ and $\PRHopcroftEqRel{(\nondet_\Sigma, \lift{\nondet_{\Sigma}}), \wPred}$
      split $C_{\epsilon}$ and obtain $C_0$ and $C_1$.
      The $k_0$ chosen at Line~\ref{line:choose-an-index} is $0$ (i.e. $C_0$) in both algorithms,
      and predecessors of $z$ are marked as dirty (the centre figure of \cref{fig:execution-example}).
  
      In the second iteration, both $\PRHopcroftEqRel{(\nondet_\Sigma, \lift{\nondet_{\Sigma}}), \wCard}$ and $\PRHopcroftEqRel{(\nondet_\Sigma, \lift{\nondet_{\Sigma}}), \wPred}$
      split $C_0$ and obtain $C_{00}$ and $C_{01}$.
      The $k_0$ chosen at Line~\ref{line:choose-an-index} of $\PRHopcroftEqRel{(\nondet_\Sigma, \lift{\nondet_{\Sigma}}), \wCard}$
      is $1$ (i.e. $C_{01}$) because $\wCard(C_{01}) = 3 > 1 = \wCard(C_{00})$, and
      predecessors of $w$ are marked as dirty (the left figure of \cref{fig:execution-example})..
      The $k_0$ chosen at Line~\ref{line:choose-an-index} of $\PRHopcroftEqRel{(\nondet_\Sigma, \lift{\nondet_{\Sigma}}), \wPred}$
      is $0$ (i.e. $C_{00}$) because $\wPred(C_{00}) = 4 > 3 = \wPred(C_{01})$, and
      predecessors of the states in $C_{01}$ are marked as dirty (the right figure of \cref{fig:execution-example}).
  
      The algorithm $\PRHopcroftEqRel{(\nondet_\Sigma, \lift{\nondet_{\Sigma}}), \wPred}$ marks
      states as dirty less than $\PRHopcroftEqRel{(\nondet_\Sigma, \lift{\nondet_{\Sigma}}), \wCard}$.
      Hence, $\PRHopcroftEqRel{(\nondet_\Sigma, \lift{\nondet_{\Sigma}}), \wPred}$
      terminates in fewer steps compared with $\PRHopcroftEqRel{(\nondet_\Sigma, \lift{\nondet_{\Sigma}}), \wCard}$.
  \end{myexample}
\end{toappendix}

\begin{credits}
  \subsubsection{\ackname}
  The authors are supported by ERATO HASUO Metamathematics for Systems Design Project (No. JPMJER1603), JST.
  T.\ Sanada and R.\ Kojima are supported by JST Grant Number JPMJFS2123.
  Y.\ Komorida is supported by JSPS KAKENHI Grant Number JP21J13334.
\end{credits}

%
%
%
%
\bibliographystyle{splncs04}
\bibliography{ronbun}
%
%
%
%
\end{document}